\documentclass[journal]{IEEEtran}

\usepackage{amsmath,graphicx}
\usepackage{amssymb}
\usepackage{lipsum}        
\usepackage{wrapfig}
\usepackage{hyperref}

\usepackage{amsthm}
\usepackage{cleveref}
\usepackage{cite}
\usepackage[normalem]{ulem}
\usepackage{subcaption} 
\usepackage{float}
\usepackage{booktabs}
\usepackage{multirow}

\newcounter{assumption}
\newcommand{\assumption}[1]{
   \refstepcounter{assumption}
   \label{#1}
   \noindent\textbf{Assumption \theassumption:}
}
\newcommand{\dist}[1]{\mathrm{dist}\left(#1\right)}
\newcommand{\euclnorm}[1]{\left\lVert#1\right\rVert_2}

\newtheorem{theorem}{Theorem}[section]

\newtheorem{lemma}[theorem]{Lemma}
\newtheorem{corollary}[theorem]{Corollary}

\input{macros.tex}

\begin{document}

%

\title{Robust Phase Retrieval by Alternating Minimization}
%
%
%
\author{Seonho Kim,~\IEEEmembership{Student Member,~IEEE,}
        and~Kiryung Lee,~\IEEEmembership{Senior Member,~IEEE}
\thanks{Seonho Kim and Kiryung Lee are with the Department of ECE, The Ohio State University, Columbus, OH 43210 USA (e-mail: kim.7604@osu.edu). This work was supported in part by NSF CAREER Award CCF-1943201. A preliminary version of this work will be presented at the 2024 IEEE International Conference on Acoustics, Speech and Signal Processing (ICASSP) \cite{kim2024robust}} 
}

%
%

\markboth{Journal of \LaTeX\ Class Files,~Vol.~14, No.~8, August~2015}%
{Shell \MakeLowercase{\textit{et al.}}: Bare Demo of IEEEtran.cls for IEEE Journals}
%



\maketitle


%
%
\begin{abstract}
We consider a least absolute deviation (LAD) approach to the robust phase retrieval problem that aims to recover a signal from its absolute measurements corrupted with sparse noise. To solve the resulting non-convex optimization problem, we propose a robust alternating minimization (Robust-AM) derived as an unconstrained Gauss-Newton method. To solve the inner optimization arising in each step of Robust-AM, we adopt two computationally efficient methods for linear programs. We provide a non-asymptotic convergence analysis of these practical algorithms for Robust-AM under the standard Gaussian measurement assumption. 
These algorithms, when suitably initialized, are guaranteed to converge linearly to the ground truth at an order-optimal sample complexity with high probability while the support of sparse noise is arbitrarily fixed and the sparsity level is no larger than $1/4$. Additionally, through comprehensive numerical experiments on synthetic and image datasets, we show that Robust-AM outperforms existing methods for robust phase retrieval offering comparable theoretical performance guarantees.
\end{abstract}

\section{Introduction}
\label{sec:intro}


Phase retrieval refers to the recovery of unknown signals $\mb x_\star\in\mathbb{R}^d$ (or $\mathbb{C}^d$) from the magnitudes of its linear measurements, which are formulated as
\begin{equation}
\label{eq:phase_formul}
b_i=|\langle\mb a_i,\mb x_\star\rangle|,\quad i=1,\ldots,m,
\end{equation} 
where $\mb a_1,\dots, \mb a_m \in \mathbb{R}^d$ (or $\mathbb{C}^d$) and 
are known measurement vectors. 
Solving the set of nonlinear equations in \eqref{eq:phase_formul} arises in numerous applications including X-ray crystallography, diffraction and array imaging, and optics (e.g. \cite{walther1963question,bunk2007diffractive,chai2010array,shechtman2015phase}). 
We consider the robust phase retrieval from the noisy amplitude measurements in \eqref{eq:phase_formul} corrupted with sparse noise, i.e. 
\begin{equation}
\label{eq:measurements}
b_i = 
\begin{cases} 
\xi_i & \text{if } i \in I_{\mathrm{out}} \\ 
|\langle\mb a_i,\mb x_\star\rangle| & \text{if } i \in I_{\mathrm{in}}
\end{cases}
\end{equation} 
where $I_{\mathrm{out}}\subset[m]$ and $I_{\mathrm{in}}=[m]\setminus I_{\mathrm{out}}$ collect the unknown indices of outliers and inliers respectively, and $\{\xi_i\}_{i\in I_{\mathrm{out}}}$ is an arbitrary sequence in $\mathbb{R}$.
For example, such a scenario arises in phase retrieval imaging applications \cite{weller2015undersampled} due to various reasons including detection failures and recording errors.

A suite of methods designed for the plain phase retrieval \cite{dong2023phase} has been adapted to address the outliers. These methods provide not only empirically successful performances but also theoretical analyses under random measurement models. For instance, anchored regression \cite{bahmani2017phase} and PhaseMax \cite{goldstein2018phasemax} formulate phase retrieval given an initial estimate as a linear program. RobustPhaseMax \cite{hand2016corruption} modifies these methods to offer robust estimation by introducing an auxiliary variable to describe the outliers. In another example, Reshaped Wirtinger Flow (RWF) \cite{zhang2017nonconvex} and Amplitude Flow \cite{wang2017solving} follow a subgradient descent approach for a least squares estimator (LSE). Median-RWF \cite{zhang2018median} is a variant of these methods tailored to robust phase retrieval. Specifically, Median-RWF uses a truncation type of regularization that identifies and excludes outliers in each iteration by median-based thresholding on the consistency of the current estimate to the measurements. Median-RWF significantly improves the empirical performance of RobustPhaseMax by tolerating a higher fraction of outliers. However, the regularization of Median-RWF involves algorithmic parameters that have been tuned specifically for the Gaussian measurement model. 
However, it has not been discussed how to generalize the tuning parameters to other measurement models. 

{A recent work proposed an approach to robust phase retrieval in the classical robust regression framework in statistics \cite{duchi2019solving}. 
Instead of the least squares, they adopted the \emph{least absolute deviation} (LAD) \cite{bloomfield1983least} to enforce the consistency to the squared amplitude measurements with outliers.
The parameter estimation is then cast as a nonconvex optimization problem. They proposed a prox-linear method that updates the estimate iteratively through local linearization of the forward model. 
This algorithm can be viewed as a variant of the Gauss-Newton method that regularizes the updates with the proximity to the previous iterate. 
The prox-linear algorithm iteratively refines the estimate through a sequence of quadratic programs for prox-linear problems and provides comparable performance to Median-RWF. 
Importantly, the Gauss-Newton method does not involve any tuning parameter. 
However, for large-scale applications such as those in astronomical or medical imaging, further acceleration of this iterative method is desired. 
They developed the proximal operator graph splitting (POGS) solver for this purpose. 
}

\begin{figure}[ht]
  \centering
  \subfloat[zero]{
    \includegraphics[width=0.45\columnwidth]{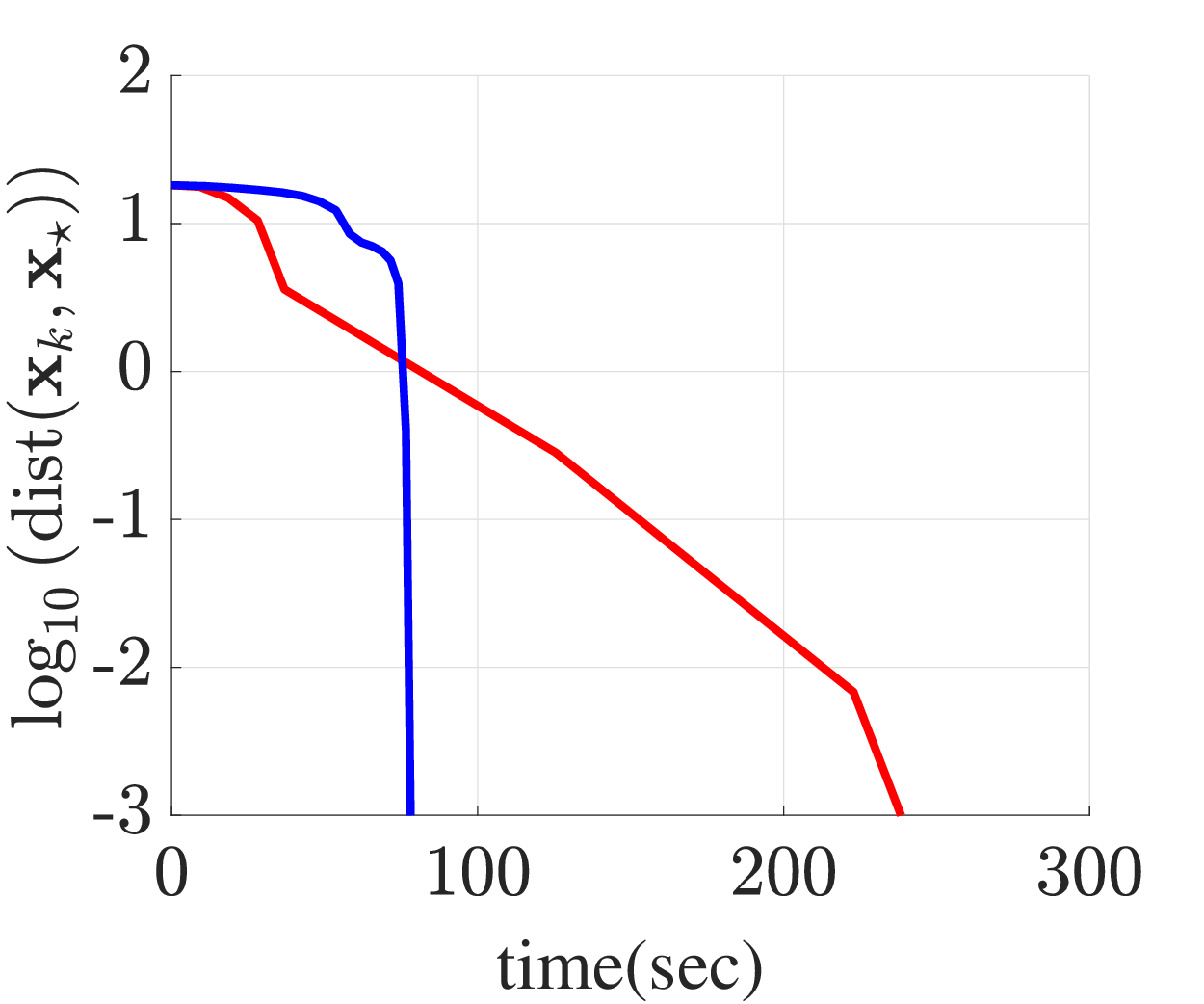}
  }
  \hfill 
  \subfloat[Cauchy distribution]{
    \includegraphics[width=0.45\columnwidth]{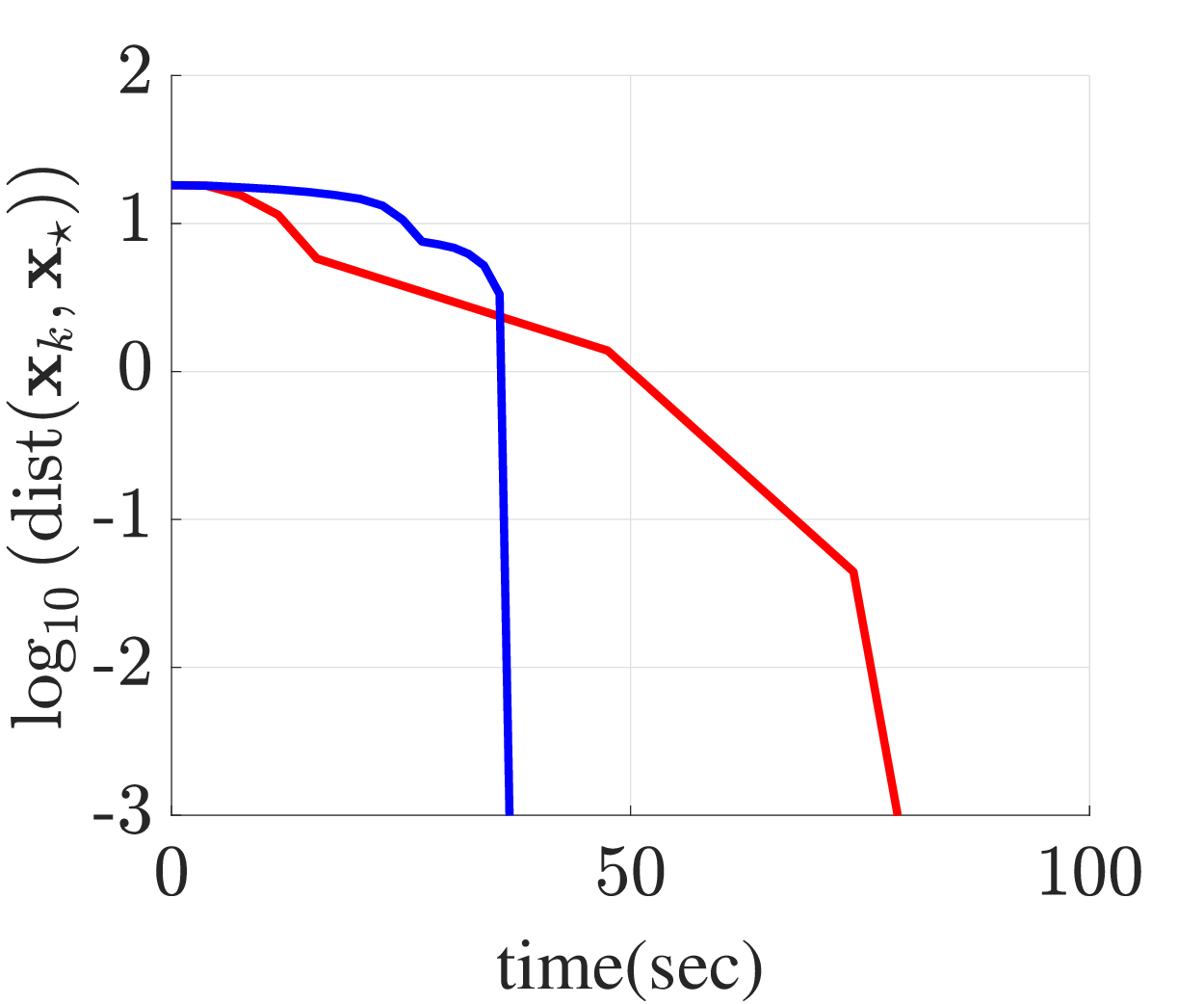}
  }
  \caption{Convergence of Robust-AM by ADMM \cite{boyd2011distributed} (blue) and prox-linear by POGS (red) in run time ($d=1,000, m=10,000,$ and $\eta=0.3$).}
  \label{fig:conv_time}
\end{figure}

In this paper, we propose an optimization approach to robust phase retrieval that shares strong theoretical guarantees (high tolerance of outlier ratio and no tuning parameters) with the prox-linear algorithm and further improves its computational cost.
The objective is achieved by a simple unconstrained Gauss-Newton method for LAD. The resulting optimization is equivalent to an alternating minimization algorithm for LAD, as described in \cite{GerchbergS72}, which is solved by a sequence of linear programs. Since this alternating minimization approach is robust in the presence of outliers, we refer to the optimization as Robust-AM.
Since this alternating minimization is a robust estimator in the presence of outliers, we refer to the optimization as Robust-AM
Our main theoretical result demonstrates that a suitably initialized Robust-AM converges to the ground-truth signal linearly from $m=\mathcal{O}(d)$ random amplitude-only measurements including up to $25\%$ outliers. 
The desired initialization can be obtained by the existing robust spectral estimators \cite{zhang2018median,duchi2019solving}. 
We verified through comprehensive numerical simulations that Robust-AM empirically outperforms the existing methods for robust phase retrieval. 
Particularly, it can tolerate a higher fraction of outliers and provide exact recovery with fewer observations. 
Furthermore, due to its unconstrained optimization formulation with the absolute amplitude measurement model, Robust-AM admits a computationally efficient ADMM algorithm, which runs faster than POGS for the prox-linear method. 
As shown in \Cref{fig:conv_time}, ADMM for Robust-AM converges faster than POGS for the prox-linear method. 
In this experiment, the fraction of outliers is set to $\eta:={|I_{\mathrm{out}}|}/{m}=0.3$, with outlier entries generated following zero and a Cauchy distribution with median $0$ and mean-absolute-deviation $1$. 
The convergence is measured by the metric \(\mathrm{dist}(\mathbf{x}, \mathbf{x}_\star) := \min_{\alpha \in \{\pm 1\}} \|\mathbf{x} - \alpha \mathbf{x}_\star\|_2\) for \(\mathbf{x},\mathbf{x}_\star\in\mathbb{R}^d\). 
\Cref{fig:conv_time} shows that the unconstrained Gauss-Newton method, without any explicit control over the proximity to previous iterates, converges to the ground truth signal \(\mathbf{x}_\star\) without overshooting.

\begin{table*}[ht!]
\centering
\caption{Comparison of RobustPhaseMax \cite{hand2016corruption}, Median-RWF \cite{zhang2018median}, Prox-linear \cite{duchi2019solving} and Robust-AM for robust phase retrieval in terms of computational cost to obtain $\epsilon$-accurate solution and sparse noise assumptions for the performance guarantees.}
\begin{tabular}{@{}l|c|c|c|c@{}} 
\toprule
Method & Computational cost & Algorithm type & Support model & Tolerable sparsity level \\ 
\midrule
\multirow{2}{*}{RobustPhaseMax} & $\mathcal{O}(m^3+(m+d)^2\log(1/\epsilon))$ & ADMM for LP \cite{wang2017new} & \multirow{2}{*}{adversarial} & \multirow{2}{*}{unspecified} \\
 &{$\tilde{\mathcal{O}}((m+d)^{2.38}\log(1/\epsilon))$} & Deterministic LP \cite{van2020deterministic} & & \\\hline
Median-RWF & $\mathcal{O}(md\log(1/\epsilon))$ & truncated gradient descent  & arbitrary fixed & unspecified \\\hline
Prox-linear & $\mathcal{O}\left(\log\log(1/\epsilon)(md^2+md\log(1/\epsilon))\right)$\footnotemark[1] & regularized Gauss-Newton (POGS) & arbitrary fixed  & $1/4$ \\\hline
Robust-AM & {$\mathcal{O}\left(m^3+(m+d)^2\log^2(1/\epsilon)\right)$} & unconstrained Gauss-Newton via \cite{wang2017new} & \multirow{2}{*}{arbitrary fixed}  & \multirow{2}{*}{$1/4$}  \\
(\Cref{thm:main}) & $\tilde{\mathcal{O}}\left((m+d)^{2.38}\log^2(1/\epsilon)\right)$  & unconstrained Gauss-Newton via \cite{van2020deterministic} & &\\
\bottomrule
\end{tabular}
\label{tab:comparison_table}

\textsuperscript{1}We establish this computational cost under the assumption that the POGS linear converges to the solution for the inner optimization of prox-linear. However, to the best of our knowledge, the convergence rate of POGS has not been shown. Thus, this computational cost is a conjecture.
\end{table*}

\noindent\textbf{Notations :} Boldface lowercase letters denote column vectors. 
We use $\|\cdot\|_1$ and  $\|\cdot\|_{2}$ to denote the $\ell_1$ norm and the Euclidean norm respectively. For brevity, the shorthand notation $[n]$ denotes the set $\{1,\ldots,n\}$ for $n\in\mathbb{N}$. We adopt the big-O notation so that $q \lesssim p$ is alternatively written as $q=\mathcal{O}(p)$. With a notation $\tilde{\mathcal{O}}$, we ignore logarithmic factors.

\section{Robust Alternating Minimization}
\label{sec:probform}



We consider the minimization of the composite function $\ell = h \circ F$ where $h: \mathbb{R}^m \to \mathbb{R}$ is a convex function and $F: \mathbb{R}^d \to \mathbb{R}^m$ is a nonlinear mapping. 
In the special case when $F$ is differentiable, Burke and Ferris \cite{burke1995gauss} proposed a constrained Gauss-Newton method where the amount of the update is upper-bounded by a threshold. 
Duchi and Ruan \cite{duchi2019solving} considered a variant where the constraint on the proximity on consecutive iterates is substituted by regularization with an additive penalty. 
We consider a more challenging case where $F$ is non-differentiable and propose an unconstrained Gauss-Newton method where the variable sequence $(\bm x_k)_{k \in \mathbb{N}\cup\{0\}}$ is iteratively updated by
\begin{equation}
\label{eq:Robust-AM}
\bm x_{k+1} \in \argmin_{\mb x\in\mathbb{R}^d} \, h(F(\bm x_k) + F'(\bm x_k) (\bm x - \bm x_k))
\end{equation}
where $F'(\bm x_k) \in \mathbb{R}^{m \times d}$ denotes the Clarke's generalized Jacobian matrix at $\bm x_k$ \cite{clarke1990optimization}.
Due to the local linear approximation of $F$ at $\bm x_k$ in \eqref{eq:Robust-AM}, $\bm x_{k+1}$ is obtained as a solution to a convex program. 
In a special case where $h: \mathbb{R}^m \to \mathbb{R}$ and $F: \mathbb{R}^d \to \mathbb{R}^m$ are respectively given by
\begin{equation}
\label{eq:def_h}
h(\bm z) = \|\bm z\|_1
\end{equation}
and 
\begin{equation}
\label{eq:def_F}
F(\mb x) = \left( \left|\langle\mb a_i,\mb x\rangle\right|-b_i \right)_{i=1}^m, 
\end{equation}
their composition reduces to 
\begin{equation}
\label{eq:LADestimator}
\ell(\mb x):=\frac{1}{m}\sum_{i=1}^{m}\left|\left|\langle\mb a_i,\mb x\rangle\right|-b_i\right|.
\end{equation} 
Then the minimization of $\ell$ corresponds to the LAD approach to robust phase retrieval with the absolute amplitude measurement model. 
Furthermore, given $h$ and $F$ as in \eqref{eq:def_h} and \eqref{eq:def_F}, the update rule in \eqref{eq:Robust-AM} is explicitly written as
\begin{equation}
\label{eq:Robust-AM_explicit}
\mb x_{k+1}\in\argmin_{\mb x\in\mathbb{R}^d}\sum_{i=1}^{m}\left|\langle\mb a_i,\mb x\rangle-\mathrm{sign}(\langle\mb a_i,\mb x_{k}\rangle)\cdot b_i\right|.
\end{equation}
The resulting algorithm \eqref{eq:Robust-AM_explicit}, derived from an unconstrained Gauss-Newton method of robust phase retrieval, is equivalent to an alternating minimization approach to the LAD formulation of robust phase retrieval when noisy measurements with a negative sign are discarded. 
An analogous alternating minimization for least-squares phase retrieval has been studied in the literature \cite{GerchbergS72,netrapalli2013phase}. 
Due to the robustness of LAD, we refer to the iterative algorithm by \eqref{eq:Robust-AM_explicit} as a robust alternating minimization (Robust-AM). 


Duchi and Ruan \cite{duchi2019solving} considered a similar robust phase retrieval with the squared amplitude measurement model via their regularized Gauss-Newton method. 


\section{Optimization Algorithms}
\label{sec:optimization}

This section discusses numerical algorithms for Robust-AM. 
First, we note that the optimization in \eqref{eq:Robust-AM_explicit} is equivalent to a linear program
\begin{equation}
\label{eq:estimatoraRobust-AM}
\arraycolsep=1.4pt\def\arraystretch{1}
\begin{array}{cl}
\displaystyle 
\mathop{\mathrm{minimize}}_{\mb x \in \mathbb{R}^d, (t_i)_{i=1}^m} & \displaystyle \langle {\mb t}, \mb 1_m \rangle \\ 
\mathrm{subject~to} & \displaystyle t_i \geq \langle\mb a_i,\mb x\rangle-\mathrm{sign}(\langle\mb a_i,\mb x_k\rangle)\cdot b_i,\\
& t_i \geq -\langle\mb a_i,\mb x\rangle+\mathrm{sign}(\langle\mb a_i,\mb x_k\rangle)\cdot b_i,\quad \forall i\in[m]\,
\end{array}
\end{equation}
where $\mb 1_m = [1,\dots,1]^\mathsf{T} \in\mathbb{R}^m$. 
There exist various computationally efficient numerical methods to solve linear programs. 
For example, the derandomized algorithm by van den Brand \cite{van2020deterministic} finds an exact solution to a linear program with $d$ variables and $m$ constraints at the cost of $\tilde{\mathcal{O}}\left((m+d)^c\right)$ multiplications where $c\approx 2.38$. 

To further accelerate the convergence of Robust-AM, we also adopt iterative numerical algorithms that provide an approximate solution to the inner optimization in \eqref{eq:Robust-AM_explicit}. 
In particular, we consider two alternating direction method of multipliers (ADMM) algorithms and a subgradient descent algorithm for the inner optimization. 
We refer to the Robust-AM with approximate solutions to the inner optimization by these ADMM algorithms as \textit{fast Robust-AM} since they provide a significantly lower computational cost for the entire convergence of Robust-AM to an $\epsilon$-accurate estimate.

\subsection{ADMM for LAD}

Given $\bm x_k$, the optimization in \eqref{eq:Robust-AM_explicit} is viewed as LAD for linear regression and one can use an ADMM algorithm for LAD \cite[Chapter~6.1]{boyd2011distributed}. 
To describe the update rule of the ADMM algorithm, we introduce shorthand notations for the sake of brevity. 
Let $\mb A\in\mathbb{R}^{m\times d}$ be a matrix whose $i$-th row is $\mb a_i^\T$ for $i \in [m]$, $\mb b:=(b_1,\ldots,b_m)\in\mathbb{R}^m$, and $\mb \Lambda_k=\mathrm{diag}(\mathrm{sign}(\langle\mb a_1,\mb x_k\rangle),\ldots,\mathrm{sign}(\langle\mb a_m,\mb x_k\rangle))$. 
By following \cite[Chapter~6.1]{boyd2011distributed} with an auxiliary variable $\mb y^t\in\mathbb{R}^d$ and dual variable $\mb \phi^t\in\mathbb{R}^m$, the update rules are given in a closed form as follows:
\begin{subequations}
\label{eq:ADMM_update}
\begin{align}
&\mb x^{t+1}=\mb A^+ \left(\mb y^t-\frac{1}{\rho}\mb \phi^t\right), \label{eq:ADMM_update_LS} \\
&\mb y^{t+1}=\mb \Lambda_k\mb b \nonumber \\
&\,+\mathrm{sign}\left(\mb A\mb x+\frac{1}{\rho}\mb \phi-\mb \Lambda_k\mb b\right)\odot\left[\left|\mb A\mb x+\frac{1}{\rho}\mb \phi-\mb \Lambda_k\mb b\right|-\frac{1}{\rho}\right]_+,\\
&\mb \phi^{t+1}=\mb \phi^t+\rho(\mb A\mb x^{t+1}-\mb y^{t+1}),
\end{align}
\end{subequations} 
where $\odot$ denotes the Hadamard product. 
The most expensive step in \eqref{eq:ADMM_update} is the least squares problem in \eqref{eq:ADMM_update_LS}. 
Since it repeats with the same $\mb A$, the pseudo inverse $\mb A^+$ of $\mb A$ can be pre-computed as $\mb A^+= (\mb A^\T\mb A)^{-1}\mb A^\T$ with cost $\mathcal{O}(d^3+d^2m)$ and be used on memory over iterations. 
For faster convergence, we adopt the varying step size strategy for $\rho$ \cite[Section~3.4.1]{boyd2011distributed}.
Importantly, $\mb A$ remains the same over the outer iteration of Robust-AM, the pseudo inverse is computed only once. 
The POGS algorithm \cite{parikh2014block} for the prox-linear \cite[Section~5]{duchi2019solving} involves a similar matrix inversion. 
However, since their matrix evolves over the outer iteration, unlike the fast Robust-AM with ADMM, it is necessary for POGS to repeat the matrix inversion. 
Recall that we wanted to adopt ADMM for the inner iteration of Robust-AM to accelerate the convergence with approximate solutions. 
Therefore, the convergence rate in the inner optimization is crucial. 
However, to the best of our knowledge, the convergence rate has not been shown for the above ADMM algorithm and the POGS algorithm. 
Below we will present another ADMM algorithm and a subgradient descent method for \eqref{eq:Robust-AM_explicit} with proven linear convergence in the next section. 
Despite their theoretical convergence results, the ADMM by \eqref{eq:ADMM_update} empirically outperformed the other methods.  
In our numerical studies, we found that the fast Robust-AM with ADMM by \eqref{eq:ADMM_update} provides faster empirical convergence than POGS (see \Cref{fig:conv_time}).

\subsection{ADMM for linear program with linear convergence}
\label{subsec:ADMMLR}
Wang and Shroff \cite{wang2017new} proposed the ADMM approach for a linear program and showed that their ADMM approach solves a linear program significantly faster than standard software such as CPLEX \cite{holmstrom2009user} and Gurobi \cite{gurobi2021gurobi}. 
Moreover, they showed the linear convergence result for their ADMM approach. 
To apply their approach to our linear program \eqref{eq:estimatoraRobust-AM}, we reformulate it into the standard form of a linear program (only with equality constraints) \cite[Equation~1]{wang2017new} by introducing $2m$ auxiliary variables $\mb u,\mb v\in\mathbb{R}^m$ as 
\begin{equation}
\label{eq:converted_LP}
\def\arraystretch{1.5}
\begin{array}{ll}
\displaystyle \mathop{\mathrm{minimize}}_{\mb w\in\mathbb{R}^{d+3m}} & \langle \mb c, \mb w \rangle \\
\mathrm{subject~to} & \displaystyle \mb B\mb w=\mb p_k,\quad \mb u,\mb s \geq \mb 0_m,
\end{array}  
\end{equation} 
where $\mb 0_{m}:=[0,\ldots,0]^\T\in\mathbb{R}^m$, $\mb 0_{m,d}:=[\mb 0_{m},\ldots,\mb 0_{m}]\in\mathbb{R}^{m\times d}$, and 
\[
\begin{aligned}
&\mb c:= [\mb 0_d;\,\mb 1_m;\,\mb 0_m;\,\mb 0_m]\in\mathbb{R}^{d+3m}\\
&\mb w:= [\mb x;\,\mb t;\,\mb u;\,\mb s]\in\mathbb{R}^{d+3m}\\
&\mb p_k:= [\mb \Lambda_k\mb b;\,\mb \Lambda_k\mb b]\in\mathbb{R}^{2m}\\
&\mathbf{B} := \begin{bmatrix} \mathbf{A} & -\mathbf{I}_m & \mathbf{0}_{m,m} & \mathbf{I}_m \\ \mathbf{A} & \mathbf{I}_m & -\mathbf{I}_m & \mathbf{0}_{m,m} \end{bmatrix} \in \mathbb{R}^{2m \times (d+3m)}.
\end{aligned}
\]
Then, by following \cite[Algorithm~1]{wang2017new}, the update rule is given as a closed form with auxiliary variable $\mb y^t=[\mb y_1^t;\,\mb y_2^t]\in\mathbb{R}^{d+3m}$ and dual variable $\mb z^t=[\mb z_1^t;\,\mb z_2^t]\in\mathbb{R}^{d+5m}$ for $\mb y_1\in\mathbb{R}^{d+m}$, $\mb y_2,\mb z_1\in\mathbb{R}^{2m},$ and $\mb z_2\in\mathbb{R}^{d+3m}$ as
 \begin{subequations}
\label{eq:ADMM_update_LP}
\begin{align}
&\mb w^{t+1}=\frac{1}{\rho}\left(\mb I+\mb B^\T\mb B\right)^{-1}\left(\mb B_1^\T\left(\mb z^t+\rho(\mb B_2\mb y^t-\bar{\mb p}_k)\right)+\mb c\right), \label{eq:ADMM_update_LP_1} \\
&\mb y^{t+1}=\mb w^{t+1}+\frac{\mb z_y^t}{\rho},\quad \mb y^{t+1}_2= [\mb y^{t+1}_2]_+,\\
&\mb z_1^{t+1}=\mb z_1^{t}+\rho\left(\mb B\mb x^{t+1}-\mb p\right),\,\,\mb z_2^{t+1}=\mb z_2^{t}+\rho(\mb w^{t+1}-\mb y^{t+1}),
\end{align}
\end{subequations} 
where
\[
\mb B_1 := \begin{bmatrix}
\mb B\\
\mb I_{d+3m}
\end{bmatrix}, \quad
\mb B_2 := \begin{bmatrix}
\mb 0_{d+2m,d+3m} \\
-\mb I_{d+3m}
\end{bmatrix}, \quad
\bar{\mb p}_k := \begin{bmatrix}
\mb p_k \\
\mb 0_{3m}
\end{bmatrix},
\]
and $[\cdot]_+$ takes the positive part of each entry of the input vector.
The most expensive step is the matrix inversion given in \eqref{eq:ADMM_update_LP_1}. 
It is calculated via the matrix-inversion lemma as
\[
(\mb I_{d+3m} + \mb B^\T \mb B)^{-1} = \mb I_{d+3m} - \mb B^\T (\mb I_{2m} + \mb B\mb B^\T)^{-1} \mb B
\]
with cost $\mathcal{O}(m^3)$. 
Since this step does not depend on previous outer iterations, one can use a pre-computed result on memory over the inner and outer iterations.  
Hence, by the linear convergence result \cite[Theorem~1]{wang2017new}, the cost for an $\epsilon_k$-accurate solution to \eqref{eq:converted_LP} is $\mathcal{O}\left(m^3+(m+d)^2\log(1/\epsilon_k)\right)$. 
However, due to more auxiliary variables in \eqref{eq:converted_LP} compared to \eqref{eq:Robust-AM_explicit}, in our numerical studies, the ADMM algorithm by \eqref{eq:ADMM_update_LP} showed slower convergence in the run time relative to the algorithm by \eqref{eq:ADMM_update}.  


\subsection{Subgradient descent for LAD}
Yang and Lin \cite{yang2018rsg} proposed a restarted subgradient (RSG) for non-smooth optimization. 
The specification of their subgradient descent to LAD in \eqref{eq:Robust-AM_explicit} is written as 
\begin{equation}
\label{eq:SGupdate}
\mb x^{t+1}=\mb x^t-\frac{\eta_t}{m}\sum_{i=1}^m\mathrm{sign}\left(\langle\mb a_i,\mb x^t\rangle-\mathrm{sign}(\langle\mb a_i,\mb x_i\rangle)\cdot b_i\right)\cdot\mb a_i,
\end{equation}
where $\eta_t > 0$ denotes a step size.
The step size remains the same for $T$ consecutive iterations and then decreases by half. 
They showed that the subsequence of iterates sampled at every $T$ indices converges at a linear rate for a sufficiently large $T$. 
Therefore, the cost for an $\epsilon$-accurate solution to \eqref{eq:Robust-AM_explicit} is $\mathcal{O}(mdT\log(1/\epsilon))$. 
However, in our numerical studies, RSG did not provide the fastest convergence in the run time compared with the other ADMM algorithms.

\section{Theoretical results}
\label{sec:mainresults}

In this section, we present the convergence analysis of the Robust-AM algorithms under the following assumptions. 
First, we adopt the standard random linear measurements and outliers with arbitrary support and adversarial values \cite{duchi2019solving}. 

\assumption{assum:meas} \label{assump:meas}
The measurement vectors $(\mb a_i)_{i=1}^m$ are independent copies of $\bm a \sim \mathrm{Normal}(\bm 0, \bm I_d)$.

\assumption{assum:outler} \label{assump:outlier} The outliers are supported on
an arbitrarily fixed set $I_{\mathrm{out}}$ with $|I_{\mathrm{out}}|=\eta m$ for $\eta\in[0,1/4]$ and their magnitudes $|\xi_i|$ can be adversarial.

Additionally, to provide the convergence analysis of the fast Robust-AM, we introduce an extra assumption that quantifies the suboptimality of solving \eqref{eq:inexact_min} by ADMM. 

\assumption{assum:inexact} \label{assump:inexact} 
There exists a bounded sequence $(\epsilon_k)_{k \in \mathbb{N}}$ such that $\mb x_k$ is an inexact minimizer up to the sub-optimality level $\epsilon_k$ for all $k \in \mathbb{N}$, i.e. 
\begin{equation}
\label{eq:inexact_min}
\begin{aligned}
& \sum_{i=1}^{m}\left|\mathrm{sign}(\langle\mb a_i,\mb x_{k}\rangle)\langle\mb a_i,\mb x_{k+1}\rangle-b_i\right| \\
& \leq 
\epsilon_k + \min_{\mb x\in\mathbb{R}^d}\sum_{i=1}^{m}\left|\mathrm{sign}(\langle\mb a_i,\mb x_{k}\rangle)\langle\mb a_i,\mb x\rangle-b_i\right|.
\end{aligned}
\end{equation}
We denote the highest sub-optimality level as $\epsilon_{\max}$, i.e. 
\[
\epsilon_{\max} := \max_{k \in \mathbb{N}} \epsilon_k.
\]

\begin{theorem}
\label{thm:main}
Suppose that Assumptions~\ref{assum:meas}, \ref{assum:outler}, and \ref{assump:inexact} hold. Then there exist absolute constants $C,c > 0$ and constants $\nu_\eta\in(0,1), \lambda_\eta>0 $ depending only on $\eta$, for which the following statement holds for all $\mb x_\star\in\mathbb{R}^d$ with probability at least $1-\exp(-cd)$: If $m \geq C d$ and 
\begin{equation}
\label{eq:initial_condition}
\max\left( 
\mathrm{dist}\left(\mb x_0,\mb x_\star\right), 
\lambda_\eta \epsilon_{\max} \right) 
\leq\sin(2/25)\|\mb x_\star\|_2, 
\end{equation}
then the sequence $\left(\mb x_k\right)_{k\in\mathbb{N} \cup \{0\}}$ by the fast Robust-AM algorithm satisfies
\begin{equation}
\label{eq:errorbound_noise}
\begin{aligned}
\mathrm{dist}\left(\mb x_k,\mb x_\star\right)\leq \nu_\eta^k\cdot\mathrm{dist}\left(\mb x_0,\mb x_\star\right)+\lambda_\eta\epsilon_{\max}
\end{aligned} 
\end{equation}
for all $k \in \mathbb{N}$, where $\mathrm{dist}(\bm x, \bm x_\star) := \min_{\alpha \in \{\pm 1\}} \|\bm x - \alpha \bm x_\star\|_2$.
\end{theorem}

{
The Robust-AM algorithm updates iterates with an exact solution to  \eqref{eq:Robust-AM_explicit}. 
Therefore, setting $\epsilon_{\max}$ to $0$ in \Cref{thm:main} provides a sufficient condition for the exact recovery of $\bm x_\star$ by Robust-AM. 
We compare the specification of \Cref{thm:main} to this scenario to the analogous results for competing methods: RobustPhaseMax \cite{hand2016corruption}, Median-RWF\cite{zhang2018median}, and prox-linear \cite{duchi2019solving}. 
\Cref{thm:main} as well as the previous results achieve the exact recovery when the number of observations $m$ exceeds a multiple of the signal dimension $d$.
Earlier theoretical results on RobustPhaseMax and Median-RWF showed that there exists an unspecified numerical constant so that the algorithms provide the exact recovery if the outlier fraction is below this constant. 
In contrast, the analyses of the prox-linear \cite{duchi2019solving} and Robust-AM (Theorem~\ref{thm:main}) demonstrate that these methods can tolerate outliers up to $1/4$ of the total observations. 
These theoretical guarantees consider different degrees of adversary for their outlier models. 
The performance guarantee of RobustPhaseMax by Hand \cite{hand2016corruption} assumed the highest adversary so that both the support and values of sparse noise are adversarial. 
The performance guarantees of Median-RWF by Zhang et al. \cite{zhang2018median} considered the same outlier model as in Assumption~\ref{assum:outler}, but they also introduced additive noise of a bounded norm in addition to sparse noise. 
Duchi and Ruan \cite{duchi2019solving} used the lowest adversary so that the support of sparse noise is random but the nonzero values of sparse noise can depend on the measurements. 
Despite providing performance guarantees under the highest adversary, as shown in \Cref{sec:experiment}, RobustPhaseMax showed significantly inferior empirical performance relative to the other methods in terms of the tolerable outlier ratio. 

\Cref{thm:main} establishes a local linear convergence of the Robust-AM algorithms. 
As discussed in \Cref{sec:probform}, Robust-AM has no explicit control over the amount of the update in each iteration unlike the constrained or regularized versions of the Gauss-Newton method \cite{burke1995gauss,duchi2019solving}. 
However, despite its simple form, Robust-AM provides the monotone decrease of the estimation error toward zero without any overshooting for robust phase retrieval in the setting of \Cref{thm:main}. 
All convergence analyses by \Cref{thm:main} and previous work \cite{zhang2018median,duchi2019solving} require an initialization within a neighborhood of the ground truth. 
The size of the basin of convergence was determined with an explicit numerical constant only in \cite{hand2016corruption} and \Cref{thm:main}. 
Various initialization methods with theoretical performance guarantees have been developed to obtain the desired initial estimate \cite{zhang2018median,duchi2019solving}. 
The sample complexity for these initialization methods does not exceed those for the subsequent estimators in order. 

Next, we discuss the computational costs for the robust estimators. 
First, RobustPhaseMax is formulated as a linear program and thus it can be exactly solved with $\tilde{\mathcal{O}}((m+d)^{2.38}\log(1/\epsilon))$ multiplications by derandomized algorithm \cite{van2020deterministic}. 
Furthermore, as we discussed in \Cref{subsec:ADMMLR}, there exists an ADMM algorithm for the linear program that costs $\mathcal{O}(m^3+(m+d)^2\log(1/\epsilon))$ for an $\epsilon$-accurate solution. 
Due to the term $\log(1/\epsilon)$, if the desired accuracy decreases in proportion to the size of the problem, it is preferable to use ADMM. Otherwise, the derandomized algorithm will be computationally efficient.
The other estimators are given as an iterative algorithm with a proven convergence rate. Therefore, we compare their computational costs to obtain an $\epsilon$-accurate solution. 
Median-RWF is a truncated gradient descent with the per-iteration cost of $\mathcal{O}(md)$. 
Since the linear convergence of Median-RWF has been established, the total cost is $\mathcal{O}(md\log(1/\epsilon))$. 
Unlike Median-RWF, the updates in prox-linear and Robust-AM involve a nontrivial inner optimization, respectively cast as a quadratic program and a linear program. 
One may use an exact solver for these sub-problems. 
For example, there exists an interior point method for quadratic programs with the cost $\mathcal{O}((m+d)^4)$ \cite{ye1989extension}. 
Since it has been shown that prox-linear converges quadratically, the total cost with this exact inner solver is $\mathcal{O}((m+d)^4)\log\log(1/\epsilon)$. 
The inner optimization in Robust-AM can be exactly solved at the cost $\tilde{\mathcal{O}}((m+d)^{2.38}\log(1/\epsilon))$ by the derandomized algorithm \cite{van2020deterministic}. 
Due to its linear convergence, the total cost of 
Robust-AM is $\tilde{\mathcal{O}}((m+d)^{2.38}\log(1/\epsilon))$. 
However, as shown in \Cref{thm:main}, the linear convergence of Robust-AM remains valid when the inner optimization problems are solved only approximately. 
The fast Robust-AM with the ADMM solver for linear programs has the per-iteration cost of $\mathcal{O}(m^3+(m+d)^2\log(1/\epsilon_{\max}))$ as shown in \Cref{sec:optimization}. 
Due to its linear convergence in \Cref{thm:main}, the total cost to obtain the $\epsilon+\lambda_\eta\epsilon_{\max}$ accuracy is $\mathcal{O}(m^3+(m+d)^2\log(1/\epsilon_{\max})\log(1/\epsilon))$.
In contrast, the convergence rate of POGS for the inner optimization in prox-linear has not been established. 
We summarize the comparison for the computational costs of algorithms in \Cref{tab:comparison_table}.

Lastly, we elaborate on the dependence of the parameters $\nu_{\eta}$ and $\lambda_\eta$ in \Cref{thm:main} on the outlier ratio $\eta$.
The linear convergence parameter $\nu_{\eta}$ in \eqref{eq:errorbound_noise} is explicitly specified as an increasing function of $\eta$ in the proof of \Cref{thm:main} and illustrated in \Cref{fig:conv_rate}. 
Therefore, smaller $\eta$ implies faster convergence. 
{The final error bound by \eqref{eq:errorbound_noise} with $k$ going to infinity is given as the amplification of the sub-optimality parameter $\epsilon_{\max}$ in the inner optimization by a factor of $\lambda_\eta$. 
First, similar to $\nu_\eta$, the parameter is also explicitly given as an increasing function of $\eta$ in the proof (see \Cref{fig:lambdaeta}). 
However, the final estimation can still be sufficiently small, as one can set the accuracy parameter to a sufficiently low value (less than $10^{-10}$) using linear program packages in readily available software such as CPLEX and Gurobi. 
Hence, the assumption on $\{\epsilon_i\}_{i=1}^k$ in \Cref{thm:main} is easily satisfied.
}}

\begin{figure}[ht]
  \centering
  \subfloat[$\nu_\eta$]{
    \includegraphics[width=0.45\columnwidth]{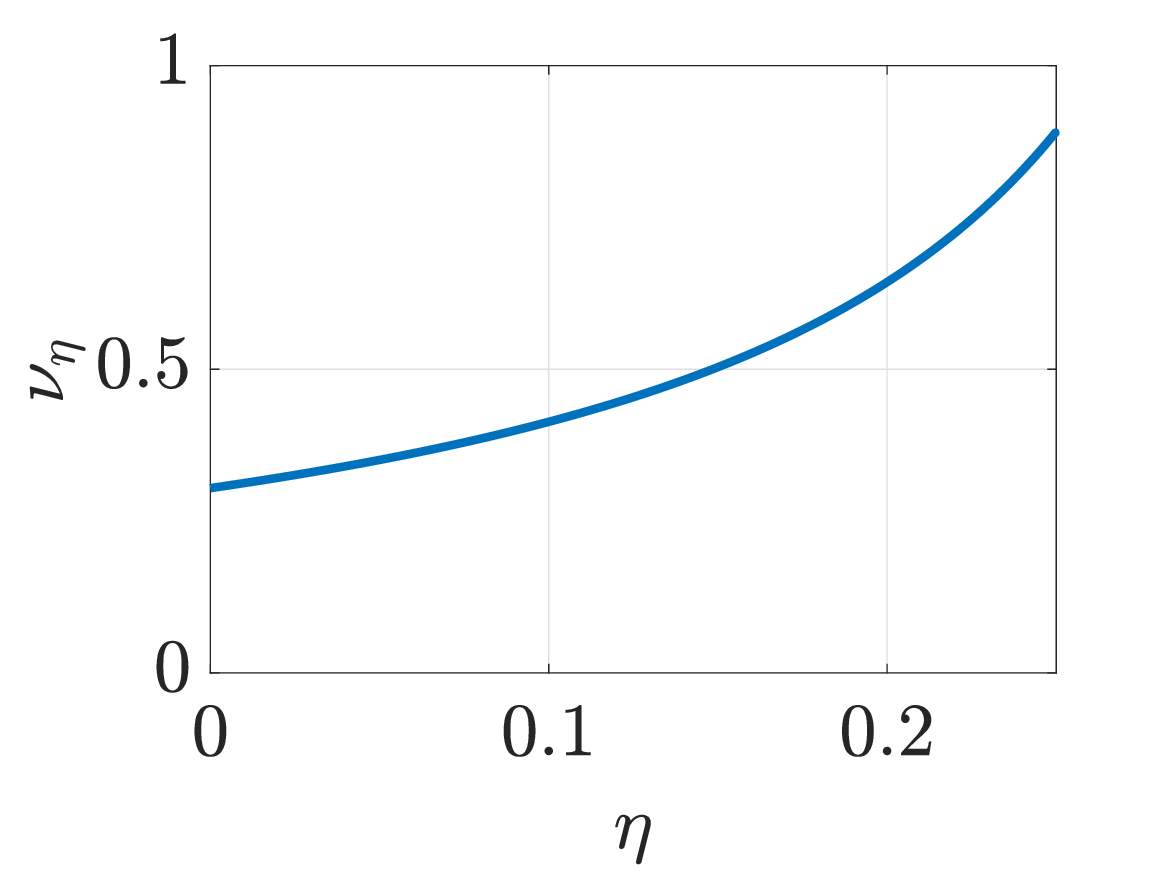}
    \label{fig:conv_rate}
  }
  \hfill 
  \subfloat[$\lambda_\eta$]{
    \includegraphics[width=0.45\columnwidth]{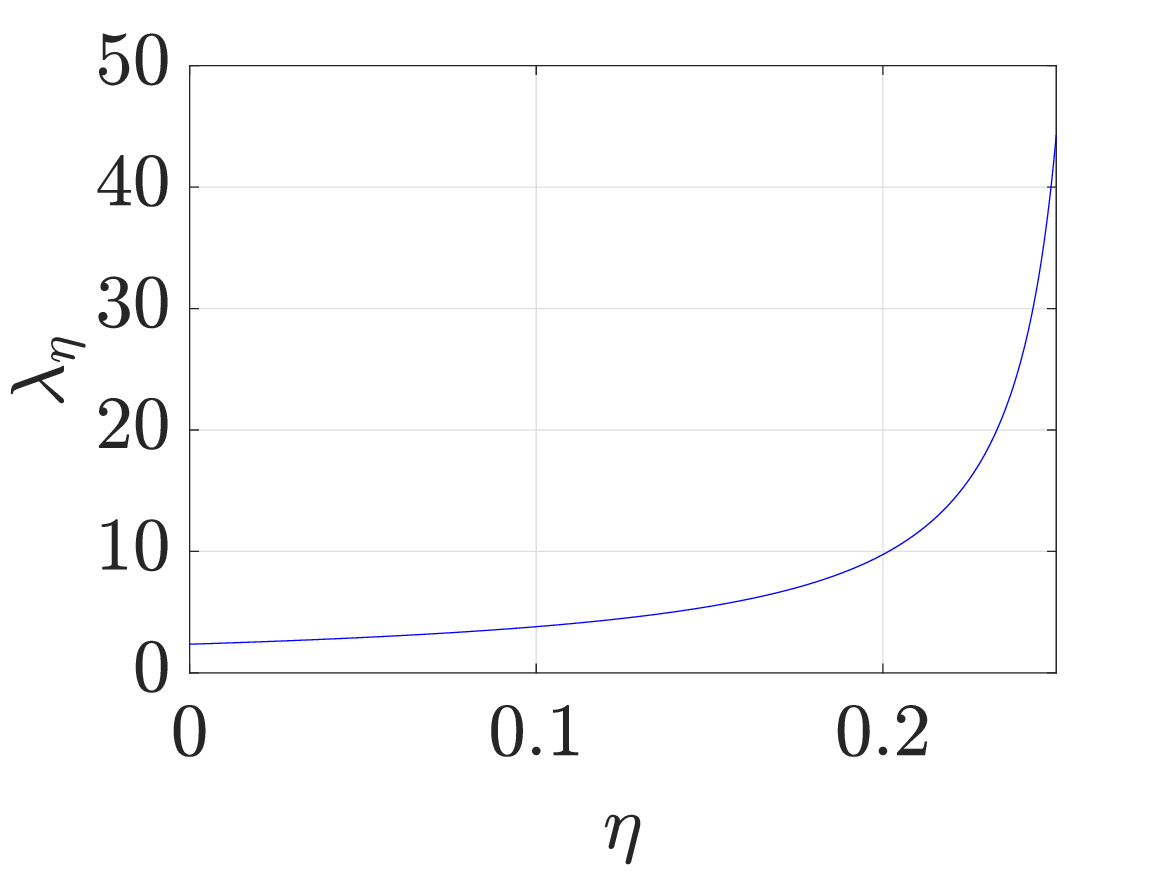}
    \label{fig:lambdaeta}
  }
  \caption{The dependence of parameters $\eta_n$ and $\lambda_n$ in \Cref{thm:main} on the outlier fraction $\eta$.}
\end{figure}

\section{Numerical Results}
\label{sec:experiment}
This section compares the empirical performances of Robust-AM to its theoretical analysis in \Cref{thm:main}. 
Robust-AM is also compared against the competing methods for robust phase retrieval, which are RobustPhaseMax, Median-RWF, and the prox-linear.
Recall that all these methods require an initial estimate. For this purpose, we adopt the spectral method by Zhang et al. \cite{zhang2018median}. 



\subsection{Synthetic data experiments}

First, through experiments on synthetic data, we show that the numerical results corroborate our theoretical findings in \Cref{thm:main} and Robust-AM outperforms the competing methods. 
In this experiment, the measurement vectors are generated so that $\{\mb a_i\}_{i=1}^m \overset{i.i.d.}{\sim} \mathrm{Normal}(\mb 0,\mb I_d)$ by following the assumptions in \Cref{thm:main} and analogous theoretical analyses of the other methods. 
The ground-truth signal is generated as $\mb x_\star\sim\mathrm{Normal}(\mb 0,\mb I_d)$ independently from the measurement vectors. 
The outlier support is randomly selected following the uniform distribution on all possible subsets $I_{\mathrm{out}} \subset [m]$ of size $\eta m$.


\begin{figure}[h]
    \centering
    \includegraphics[width=0.35\textwidth]{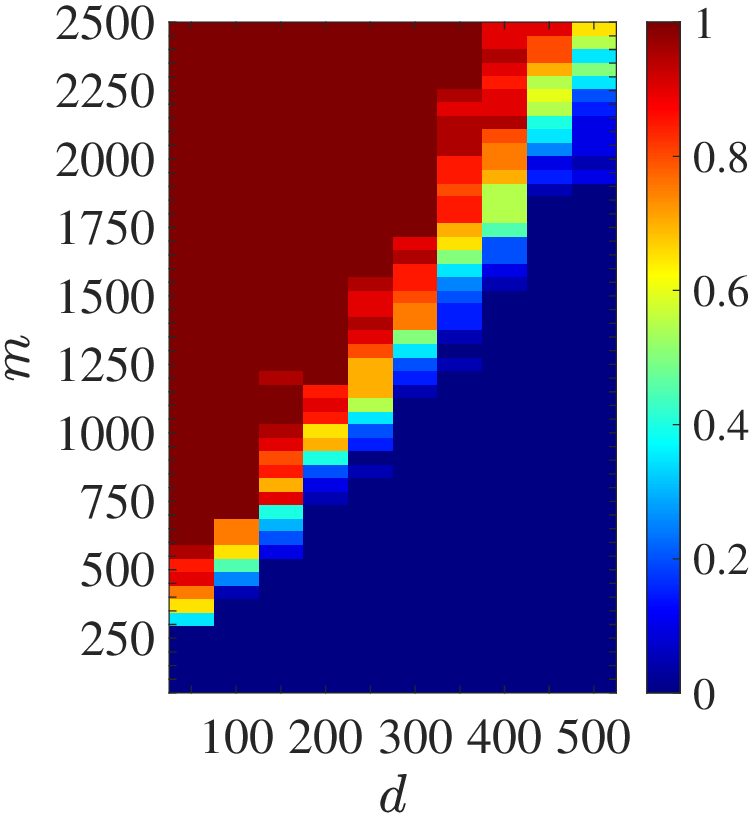}
    \caption{Phase transition of empirical success rate by Robust-AM per the number of measurements $m$ and the dimension $d$. 
    }
    \label{fig:dimension}
\end{figure}

\begin{figure}[ht!]
    \centering

      \begin{subfigure}
    {0.45\textwidth}
        \centering
        \includegraphics[width=0.45\linewidth]{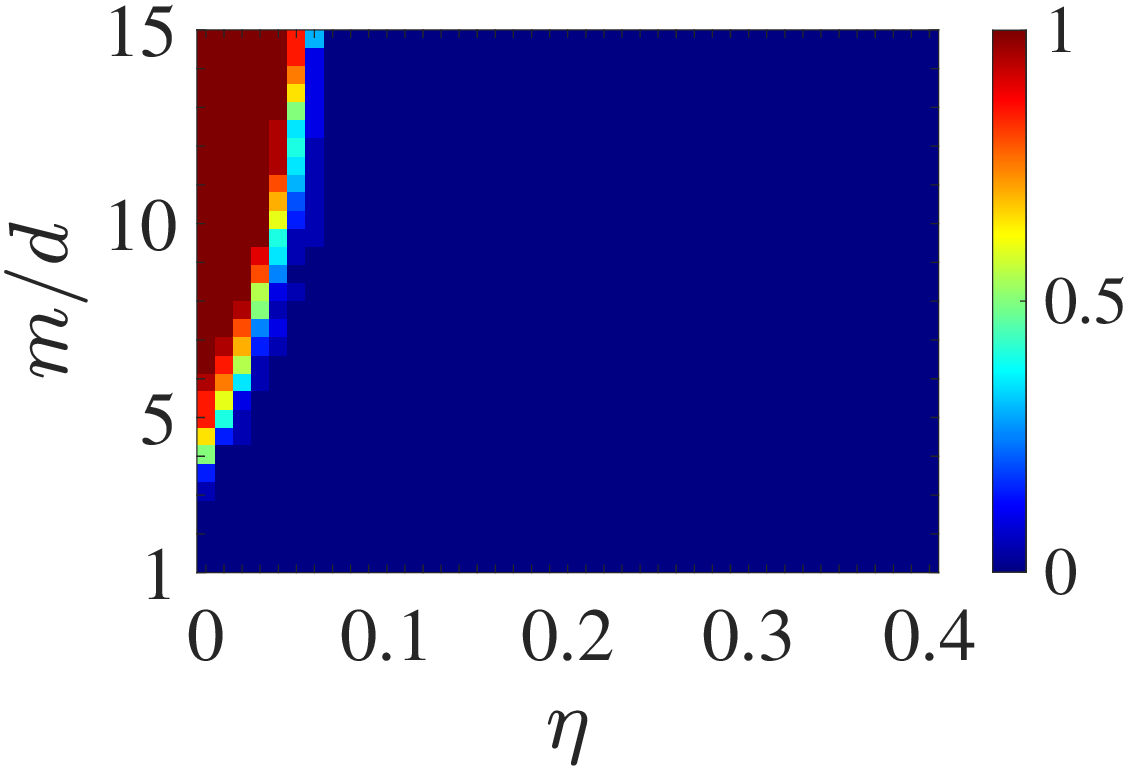}
        \hfill
        \includegraphics[width=0.45\linewidth]{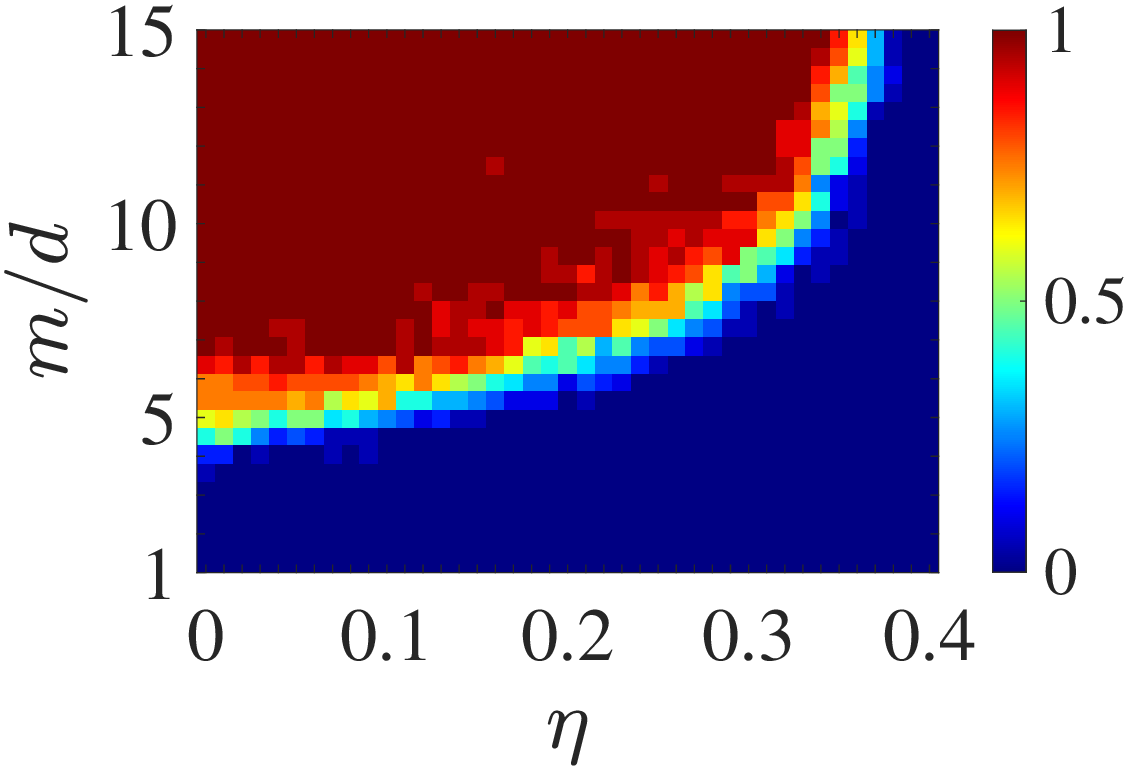}

        \vspace{0.2cm}

        \includegraphics[width=0.45\linewidth]{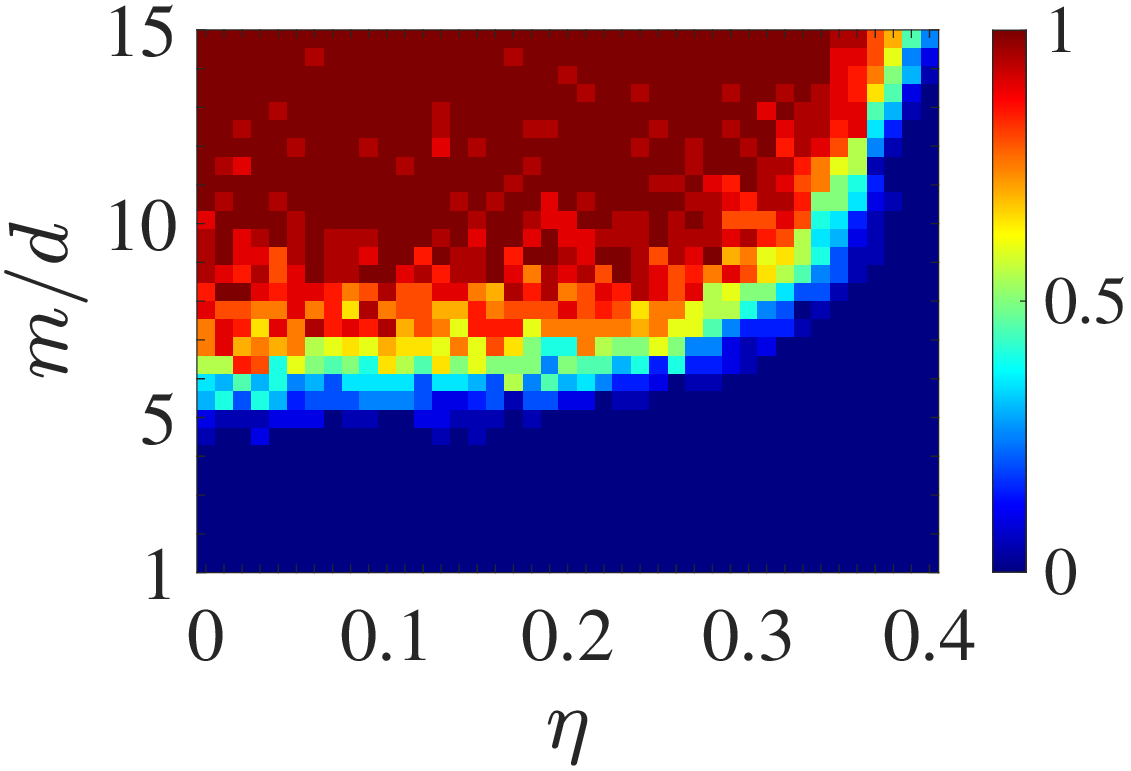}
        \hfill
        \includegraphics[width=0.45\linewidth]{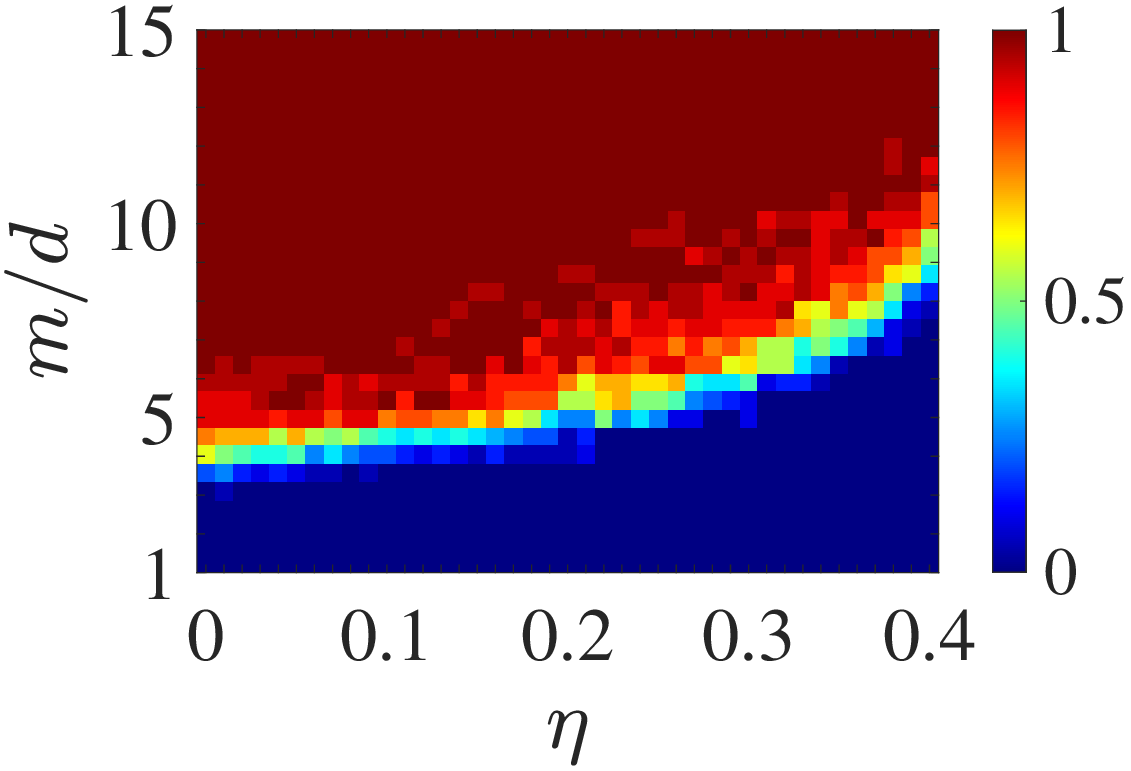}

        \caption{Cauchy distribution}
        \label{fig:Cauchy}
    \end{subfigure}

    \vspace{0.5cm}
    

    \begin{subfigure}{0.45\textwidth}      
        \centering
        \includegraphics[width=0.45\linewidth]{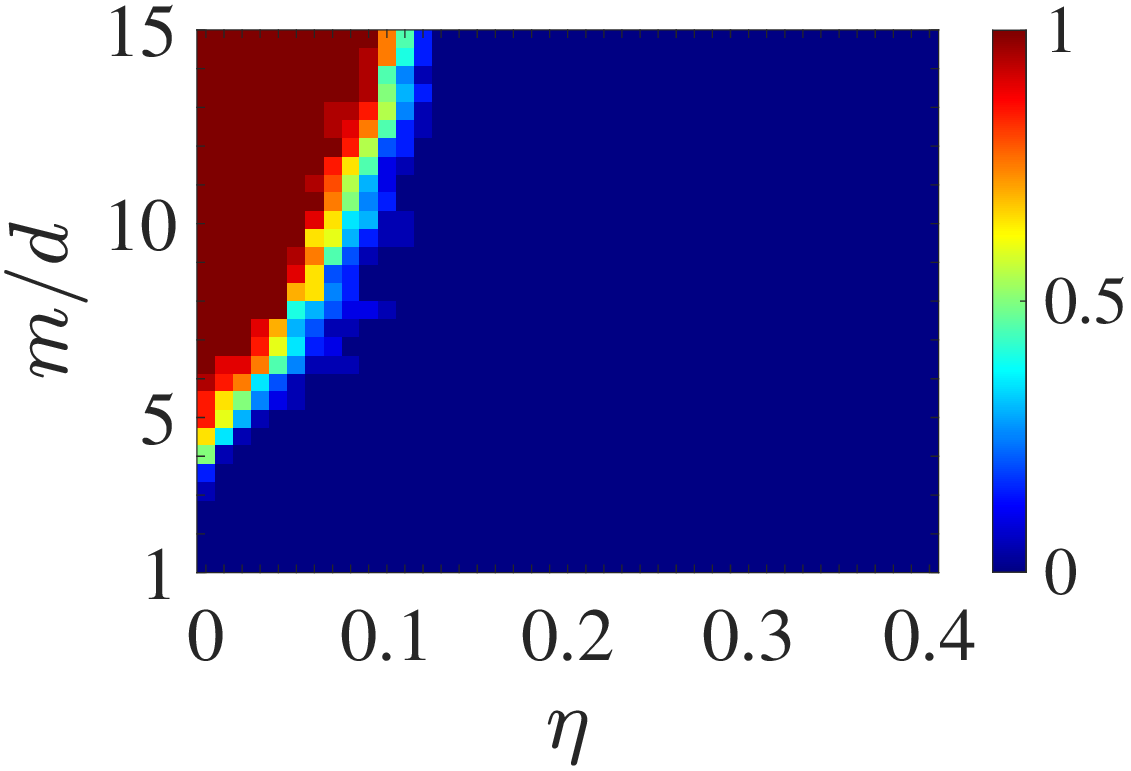}
        \hfill
        \includegraphics[width=0.45\linewidth]{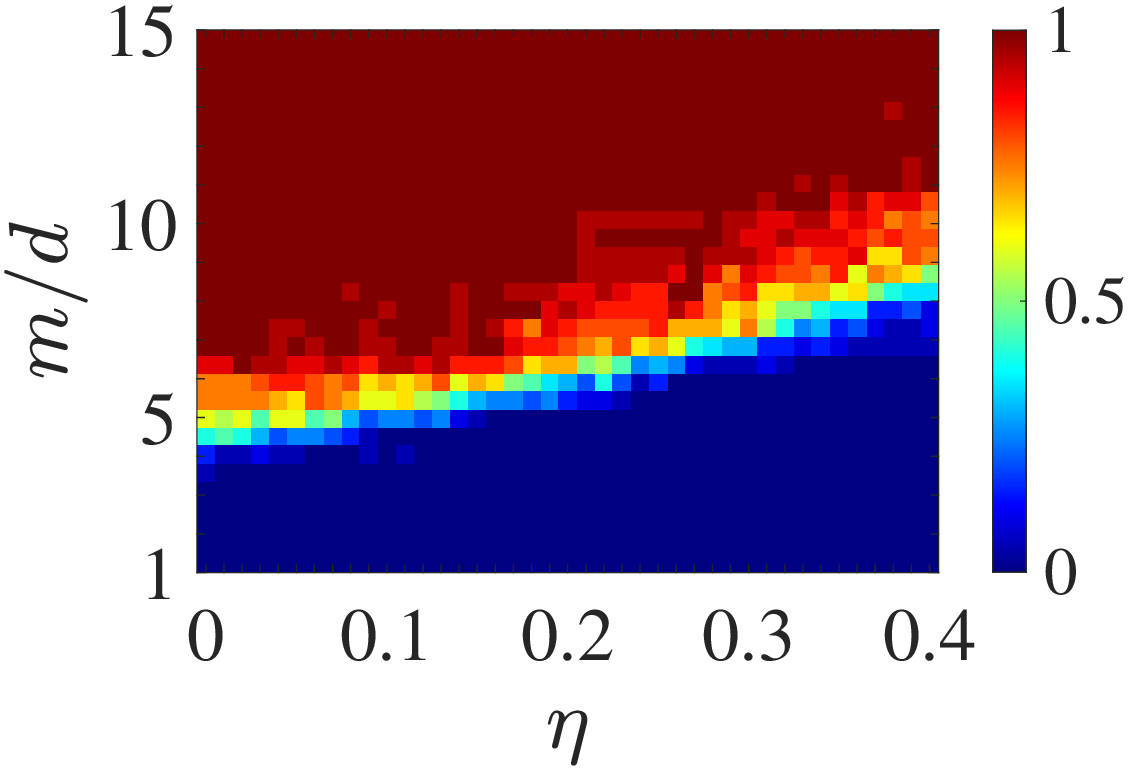}

        \vspace{0.2cm}

        \includegraphics[width=0.45\linewidth]{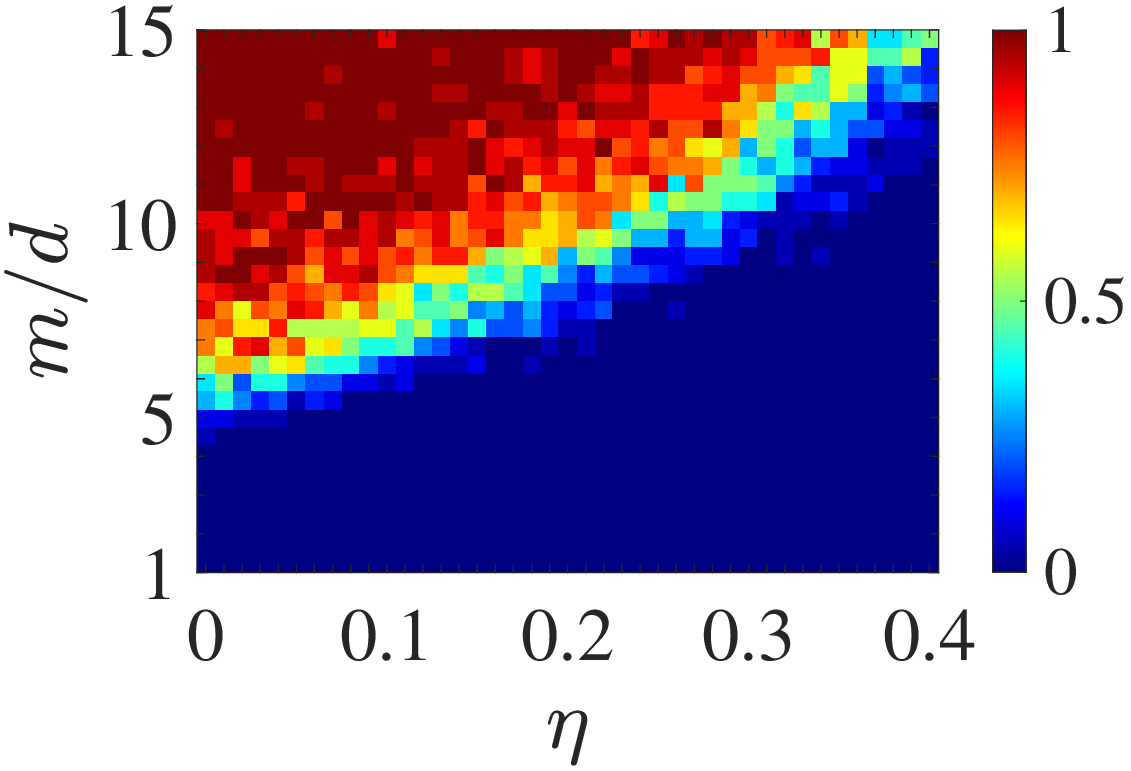}
        \hfill
        \includegraphics[width=0.45\linewidth]{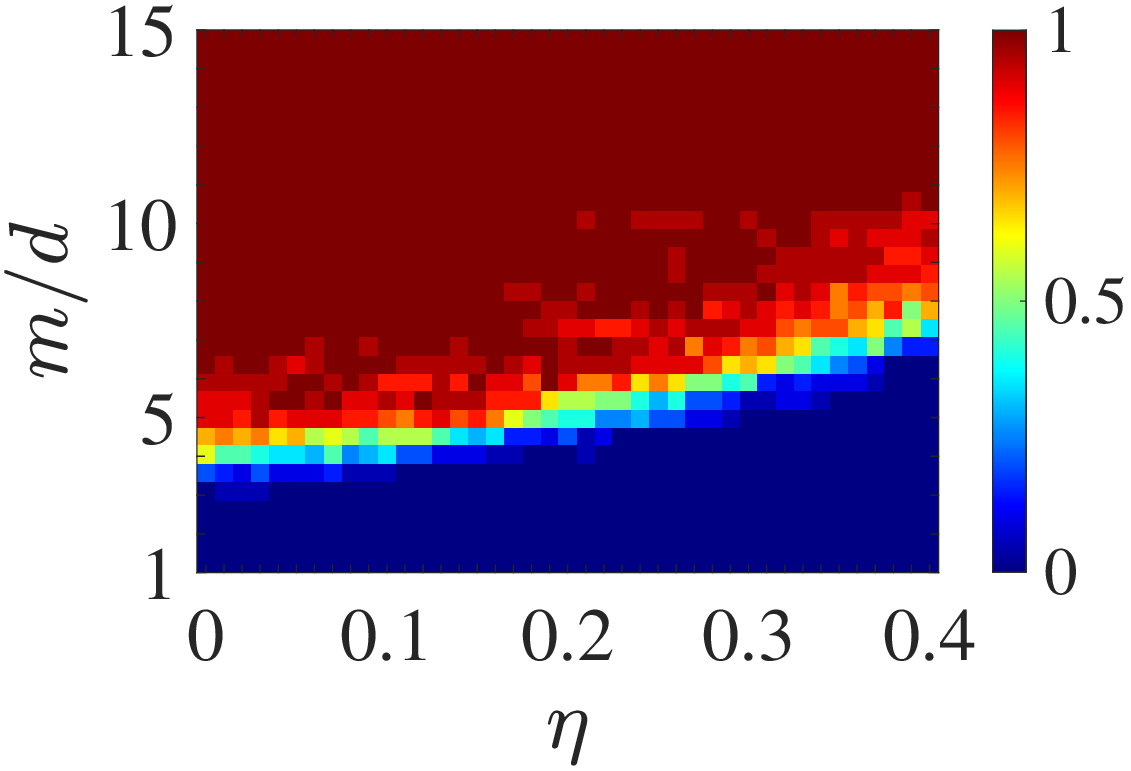}

        \caption{uniform distribution}
        \label{fig:uniform}
    \end{subfigure}

    \vspace{0.5cm}
    
    \begin{subfigure}
    {0.45\textwidth}
        \centering
        \includegraphics[width=0.45\linewidth]{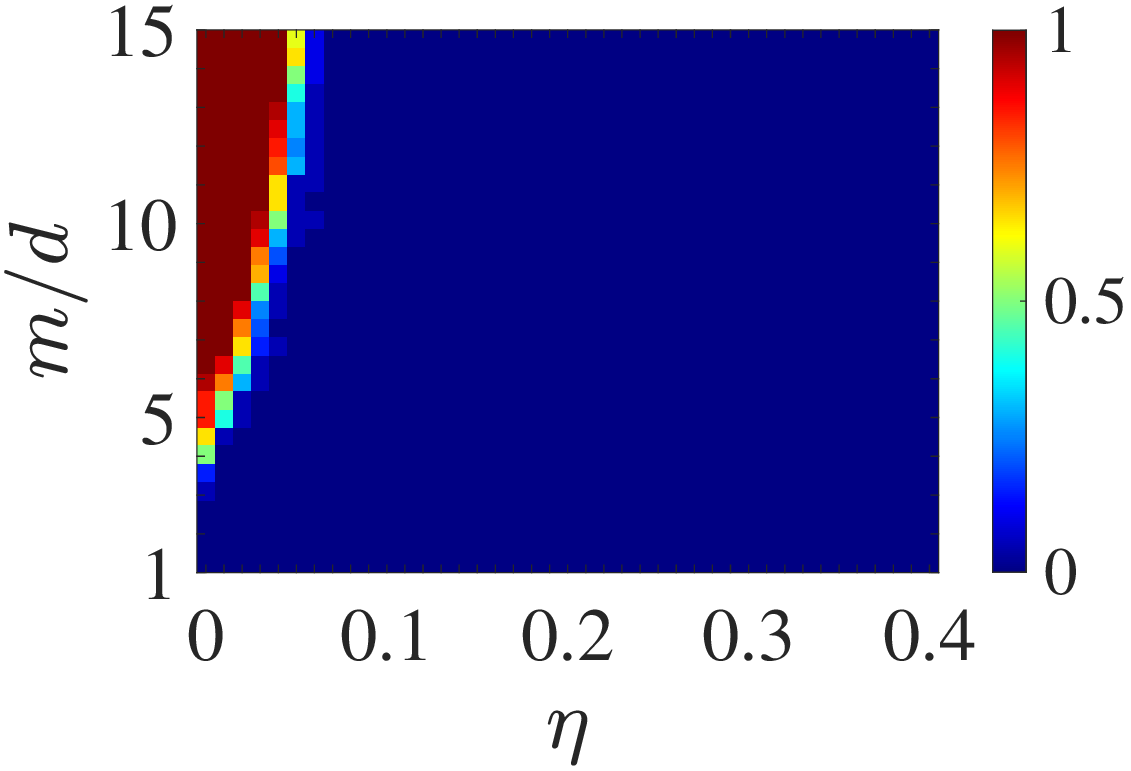}
        \hfill
        \includegraphics[width=0.45\linewidth]{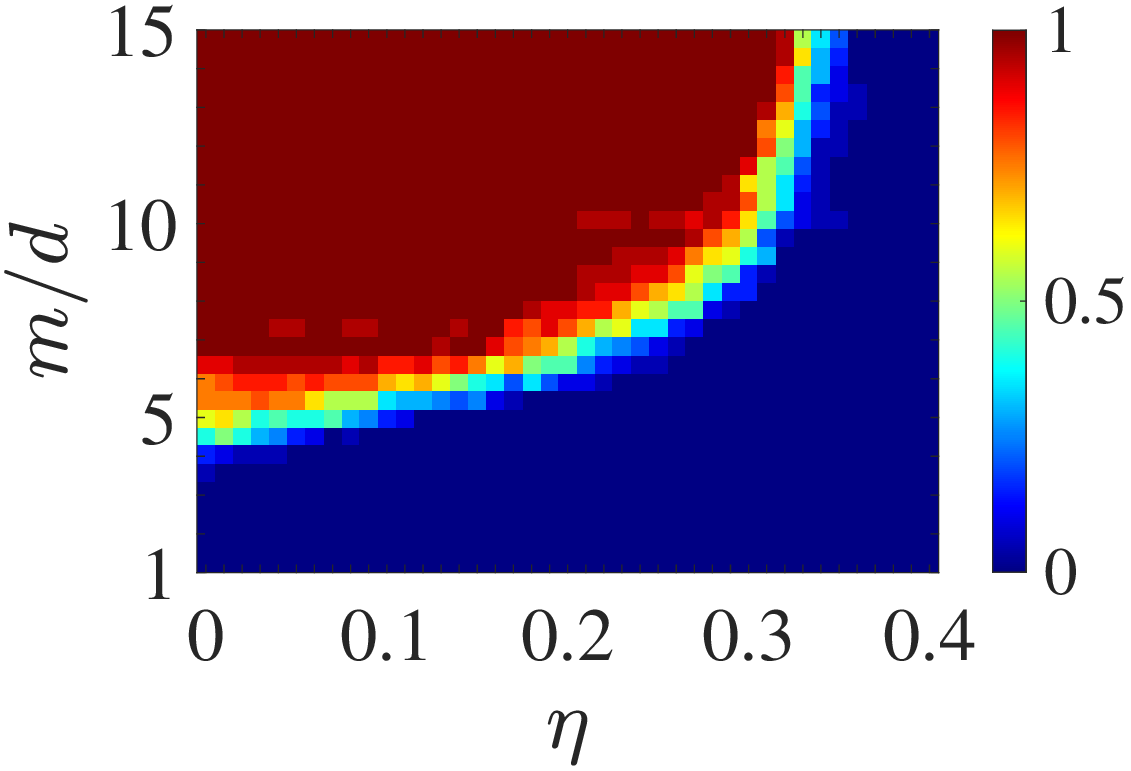}

        \vspace{0.2cm}

        \includegraphics[width=0.45\linewidth]{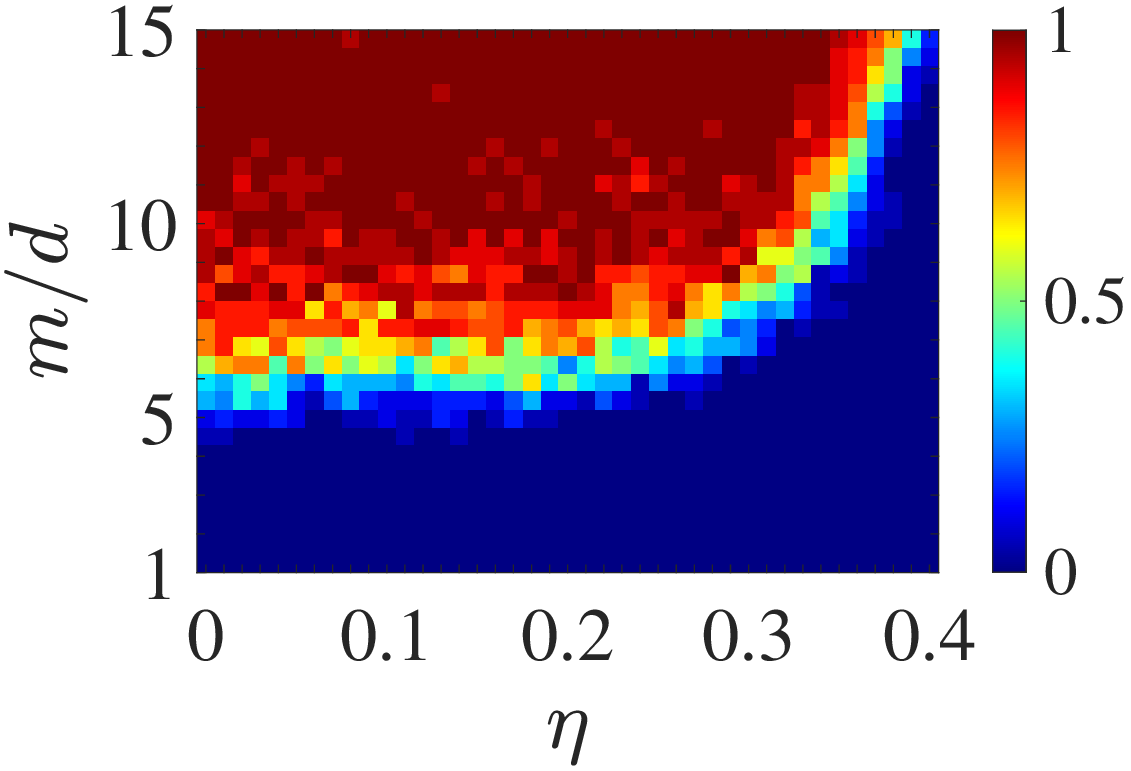}
        \hfill
        \includegraphics[width=0.45\linewidth]{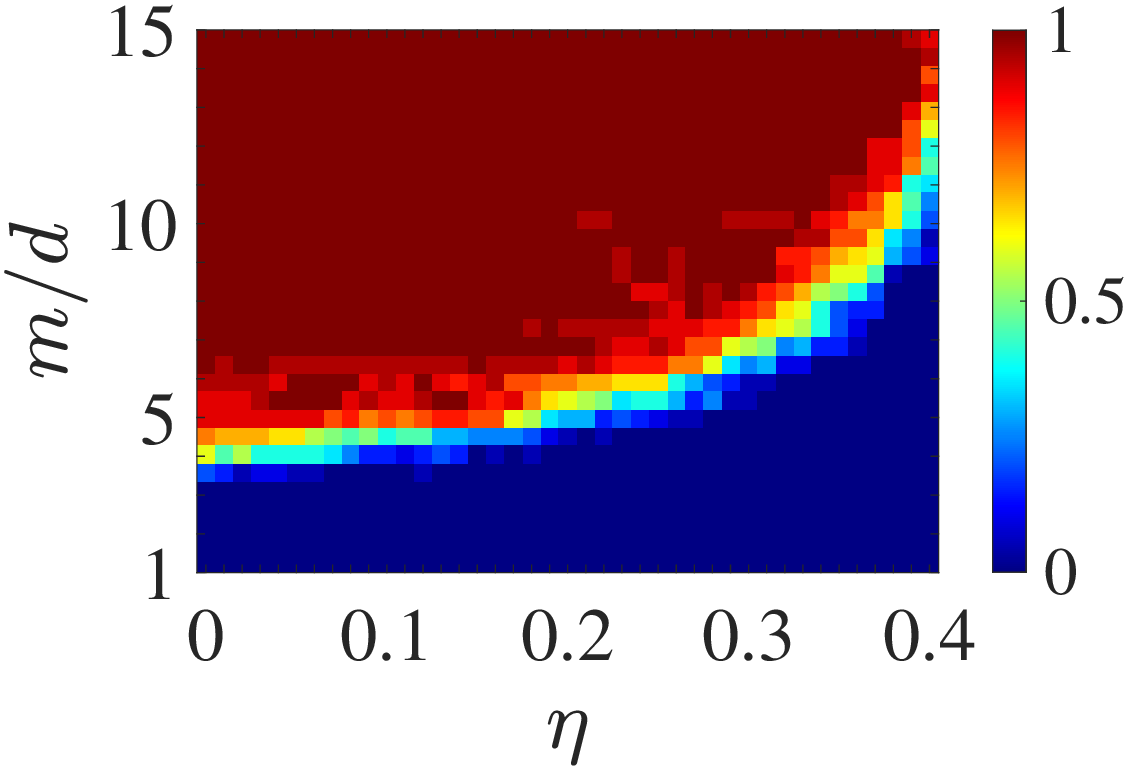}
        
        \caption{zero}
        \label{fig:zero}
    \end{subfigure}

    \caption{Phase transition of success rate per the measurement ratio $m/d$ and the fraction of outliers $\eta$ for various outlier magnitude models. Subfigures are displayed according to RobustPhaseMax (top-left), Median-RWF (top-right), prox-linear method (bottom-left), and Robust-AM (bottom-right).}
    \label{fig:first}
\end{figure}

\Cref{fig:dimension} shows the phase transition of the empirical success rate by Robust-AM through Monte Carlo simulations, where the outlier values are i.i.d. following the Cauchy distribution with median $0$ and mean-absolute-deviation $1$. 
The fraction of outliers is fixed to $\eta=0.25$
Recall that the performance guarantee in \Cref{thm:main} applies uniformly to all ground-truth signals. 
To observe the empirical performance in an analogous setting, we design the experiment as follows: 
1) Generate $20$ sets of random measurement vectors $\{\mb a_i\}_{i=1}^m$. Generate $30$ sets of random ground-truth $\mb{x}_\star$; 
2) For each fixed $\{\mb a_i\}_{i=1}^m$, success is declared if the estimator recovers all $30$ ground-truth signals by satisfying $\mathrm{dist}(\hat{\mb{x}}, \mb{x}_\star) \leq 10^{-3}$ where $\hat{\mb{x}}$ denotes the estimate;
3) The empirical success rate is calculated on the outcomes from $20$ distinct sets of measurement vectors.
The transition occurs at the boundary where the number of measurements is proportional to the ambient dimension (signal length). 
This empirical result corroborates our theoretical finding in \Cref{thm:main}. 
Next, we repeat the same experiment on RobustPhaseMax, Median-RWF, and the prox-linear. 
\Cref{fig:Cauchy} compares the empirical performance of Robust-AM against RobustPhaseMax, Median-RWF, and the prox-linear by displaying the phase transition of these methods for a range of the outlier fraction $\eta$ in this setting.
The ambient dimension is set to $d = 100$. 
\Cref{fig:Cauchy} shows that Robust-AM outperforms all the other methods with a significantly lower threshold for the phase transition.
We further expand the comparison to other models for outlier values. 
The second scenario draws $\xi_i$ from the uniform distribution on $(-d\|\mb x_\star\|_2^2/2,d\|\mb x_\star\|_2^2/2)$.
The third scenario sets $\xi_i$ to $0$. 
As observed in \Cref{fig:uniform,fig:zero}, similar trends appear in the other outlier models. 
RobustPhaseMax, while providing the strongest theoretical performance guarantee, shows the worst empirical performance in the comparison. 
There is no consistent dominance between Median-RWF and the prox-linear algorithm. 
Median-RWF outperforms the prox-linear in the second scenario, but the other way around in the other scenarios. 



\begin{figure*}[h!]
    \centering
    \begin{subfigure}[b]{0.3\linewidth}
        \includegraphics[width=\linewidth]{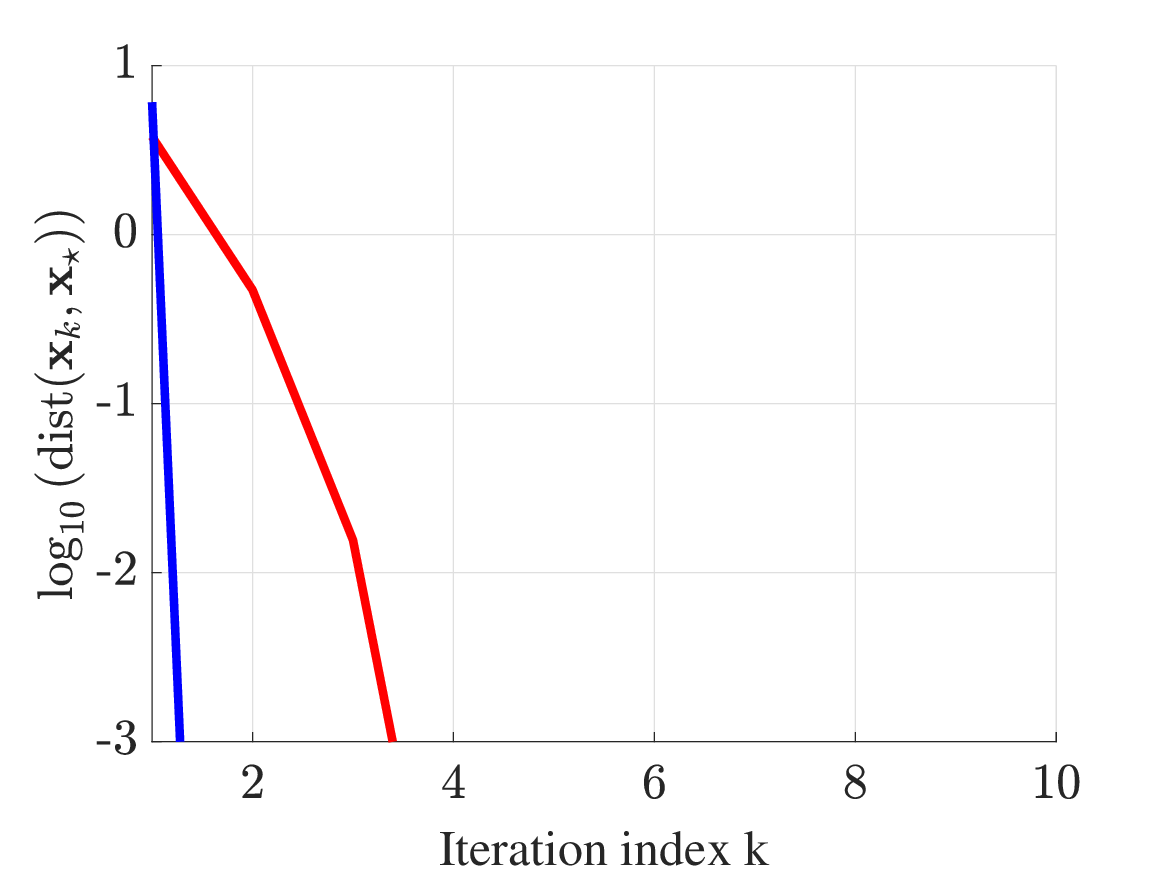}
        \caption{$\eta=0.1$}
        \label{fig:convergence1}
    \end{subfigure}
    \hfill 
    \begin{subfigure}[b]{0.3\linewidth}
        \includegraphics[width=\linewidth]{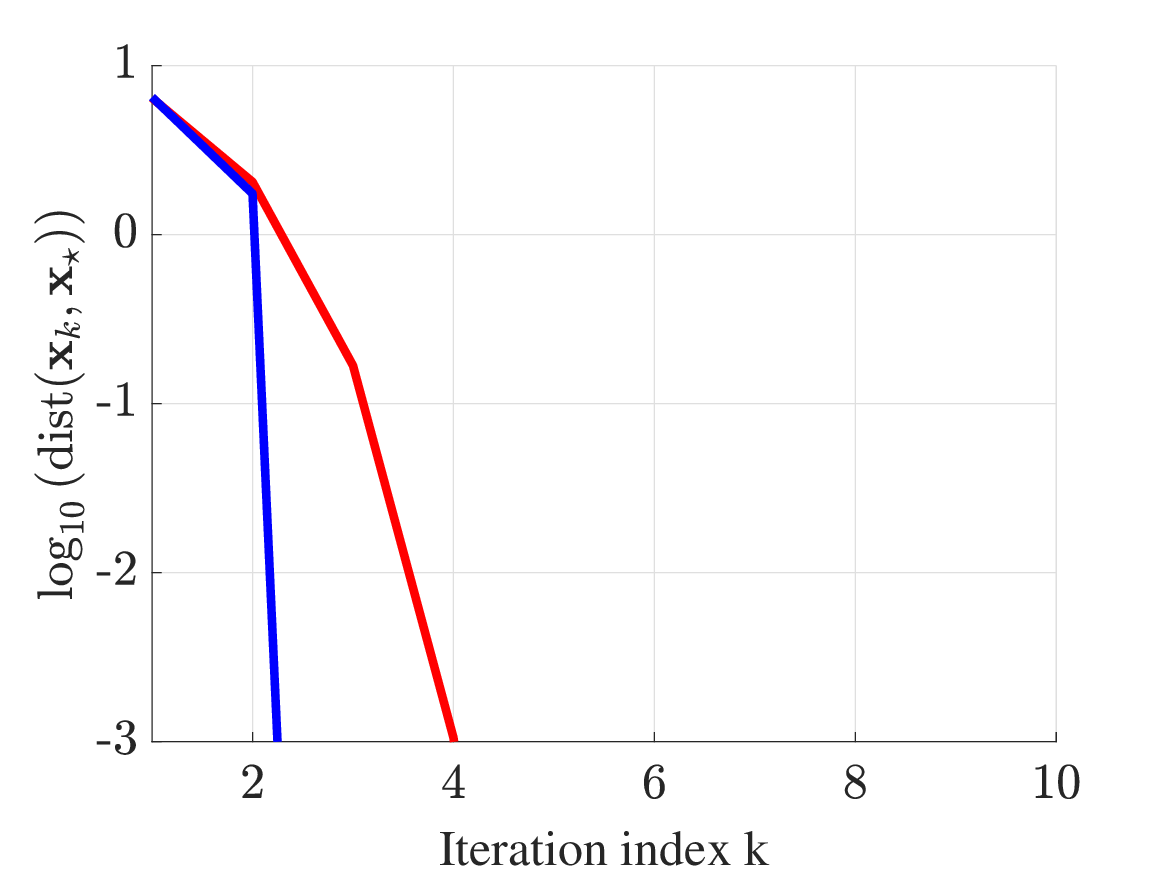}
        \caption{$\eta=0.2$}
        \label{fig:algorithm2}
    \end{subfigure}
    \hfill 
    \begin{subfigure}[b]{0.3\linewidth}
        \includegraphics[width=\linewidth]{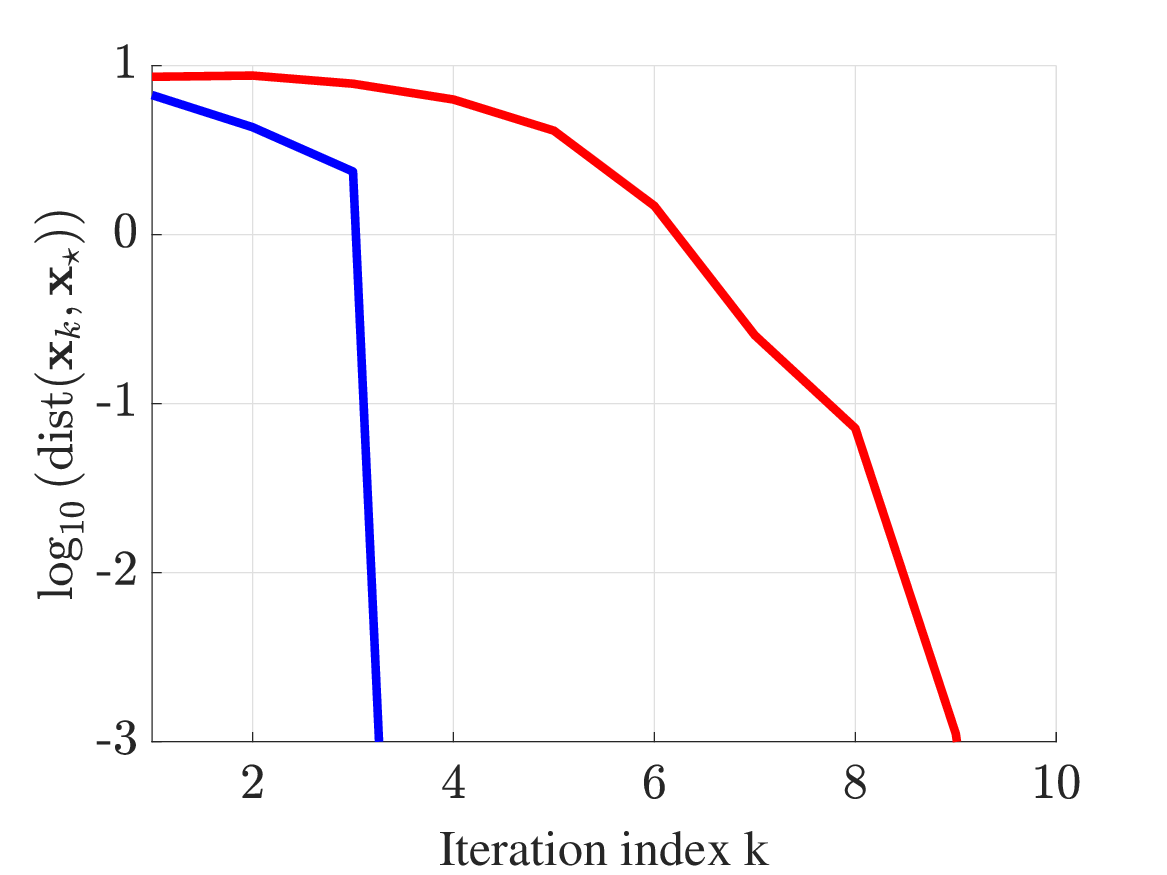}
        \caption{$\eta=0.3$}
        \label{fig:algorithm3}
    \end{subfigure}
    \caption{Convergence of Robust-AM (blue) and the prox-linear (red) in the iteration count.}
    \label{fig:convergence}
\end{figure*}




Next, we compare the convergence speed of Robust-AM and the prox-linear algorithm. 
In this experiment, the dimension parameters are set to $m=1,500$ and $d=200$ where the values of outliers are zero. The outlier ratio varies over $\eta\in\{0.1,0.2,0.3\}$.
\Cref{fig:convergence} illustrates how the log of $\mathrm{dist}({\mb x_k},\mb x_\star)$ decays over the iteration index $k$. 
The median over $10$ trials is plotted. 
In their theoretical analyses, the prox-linear algorithm converges faster at a quadratic rate than the linear convergence of Robust-AM in \Cref{thm:main}. 
However, as shown in \Cref{fig:convergence}, Robust-AM empirically converges faster than the prox-linear algorithm in the iteration count for all considered $\eta$.
Moreover, \Cref{fig:convergence} illustrates that the number of iterations for Robust-AM increases as $\eta$ increases. 
This implies that for each iteration, the convergence rate of Robust-AM is proportional to $\eta$.
This supports our theoretical finding that the convergence parameter $\nu_\eta$ in \Cref{thm:main} is an increasing function of $\eta$ as shown in \Cref{fig:conv_rate}.





\subsection{Real image experiments}


\begin{figure*}[ht!]
    \centering
    \begin{subfigure}[b]{0.22\textwidth}
        \includegraphics[width=\textwidth]{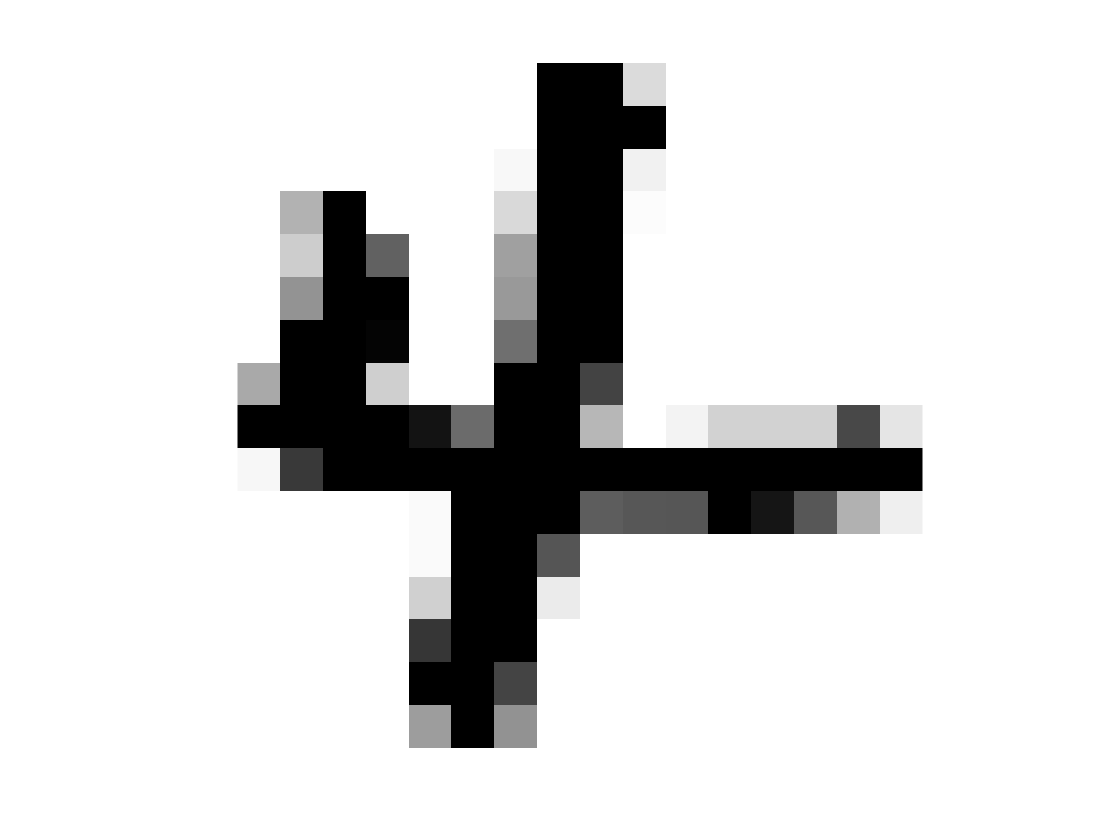}
        \caption{Ground-truth}
        \label{fig:sub1}
    \end{subfigure}
    \hfill 
    \begin{subfigure}[b]{0.22\textwidth}
        \includegraphics[width=\textwidth]{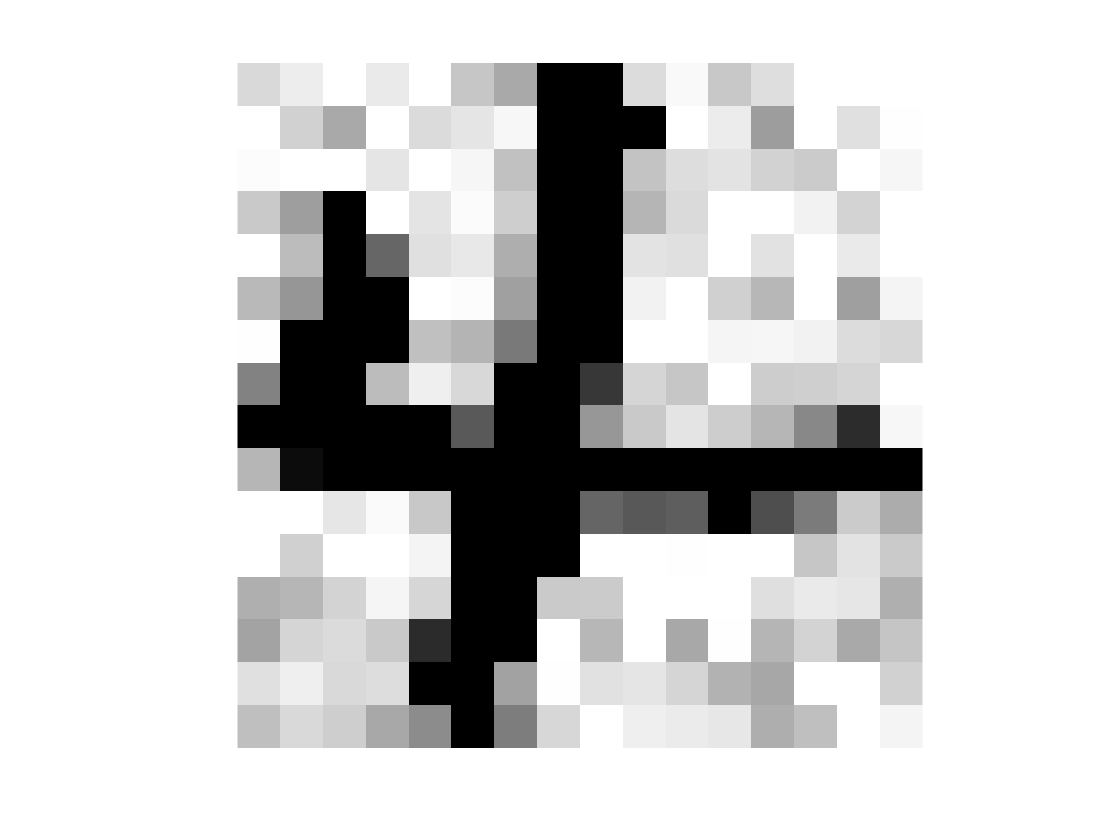}
        \caption{Recovered image by the prox-linear method}
        \label{fig:sub3}
    \end{subfigure}
    \hfill 
    \begin{subfigure}[b]{0.22\textwidth}
        \includegraphics[width=\textwidth]{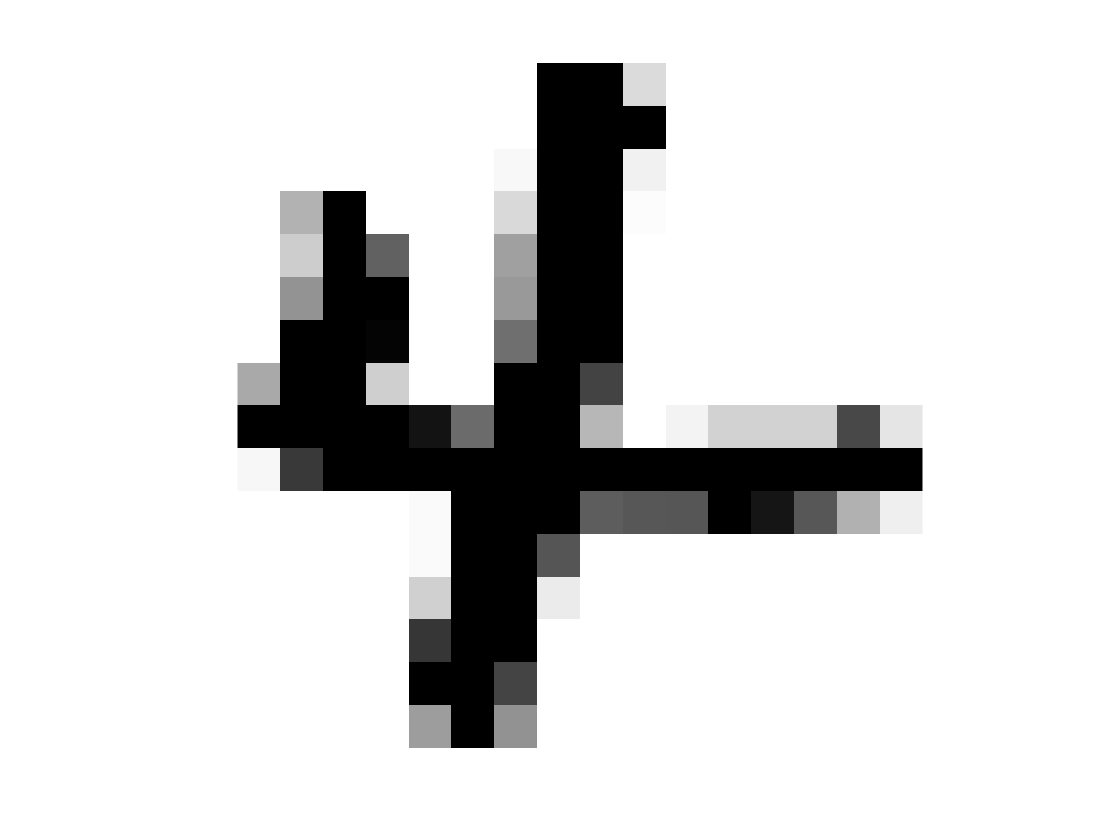}
        \caption{Recovered image by our method}
        \label{fig:sub4}
    \end{subfigure}
    \caption{Example of recovery for an image data.}
    \label{fig:process_real}
\end{figure*}

{
We further apply Robust-AM to a set of image data to show that Robust-AM continues outperforming the other competing methods for non-Gaussian measurement models.
We adopt the structured random measurement model in the experimental setting in \cite[Section~6.3]{duchi2019solving} given by
\begin{equation}
\label{eq:structured_A}
\mb A_{\mathrm{H}} = (\mb I_k \otimes \mb H_n) [\mb S_1, \mb S_2, \cdots, \mb S_k]^\T \in\mathbb{R}^{kn\times n},
\end{equation}
where $\mb H_n \in \mathbb{R}^{n\times n}$ denotes the normalized Hadamard matrix and $\mb S_1,\ldots \mb S_k\in\mathbb{R}^{n\times n}$ are diagonal matrices whose diagonal entries are independently drawn uniformly random from $\{\pm 1\}$. 
The measurement vector $\mb a_i$ is the $i$-th column of $\mb A_{\mathrm{H}}^\T$ for $i \in [m]$, where $m = kn$. 
The linear measurement operator in \eqref{eq:structured_A} applies to the vectorized version of a 2D input image $\mb X_\star\in\mathbb{R}^{n_1\times n_2}$ denoted by $\mb x_\star := \mathrm{Vec}(\mb X_\star)\in\mathbb{R}^{n}$ with $n=n_1\times n_2$. 
The measurements corresponding to outliers are substituted by zero in the experiment. 

\begin{figure}[h]
    \centering

      \begin{subfigure}
    {0.45\textwidth}
        \centering
        \includegraphics[width=0.45\linewidth]{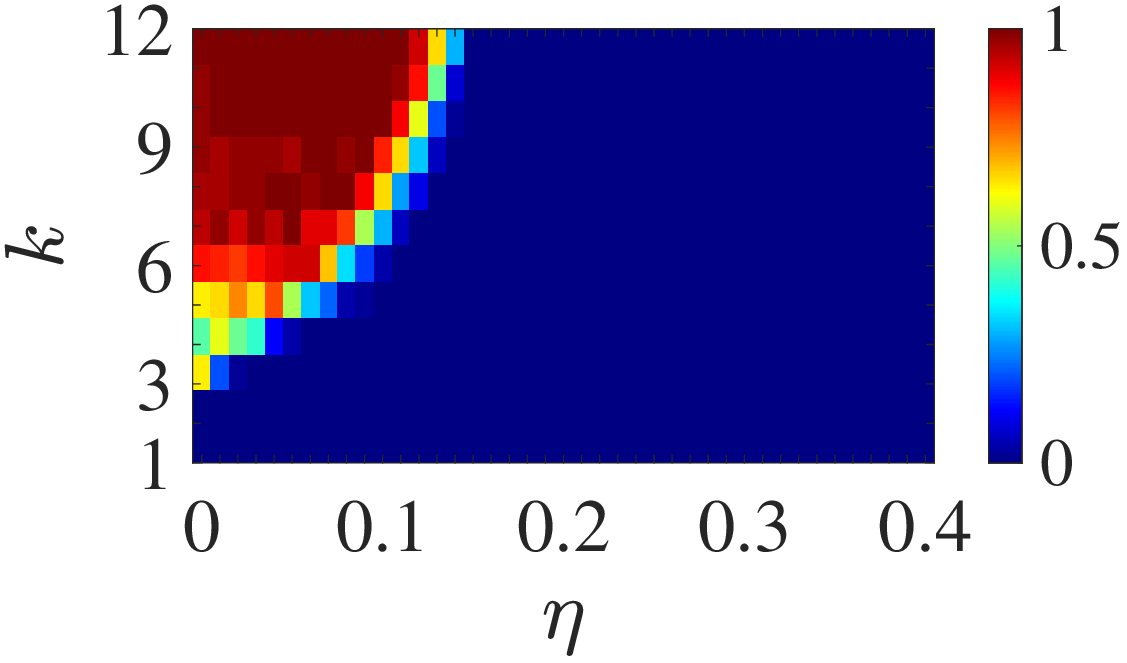}
        \hfill
        \includegraphics[width=0.45\linewidth]{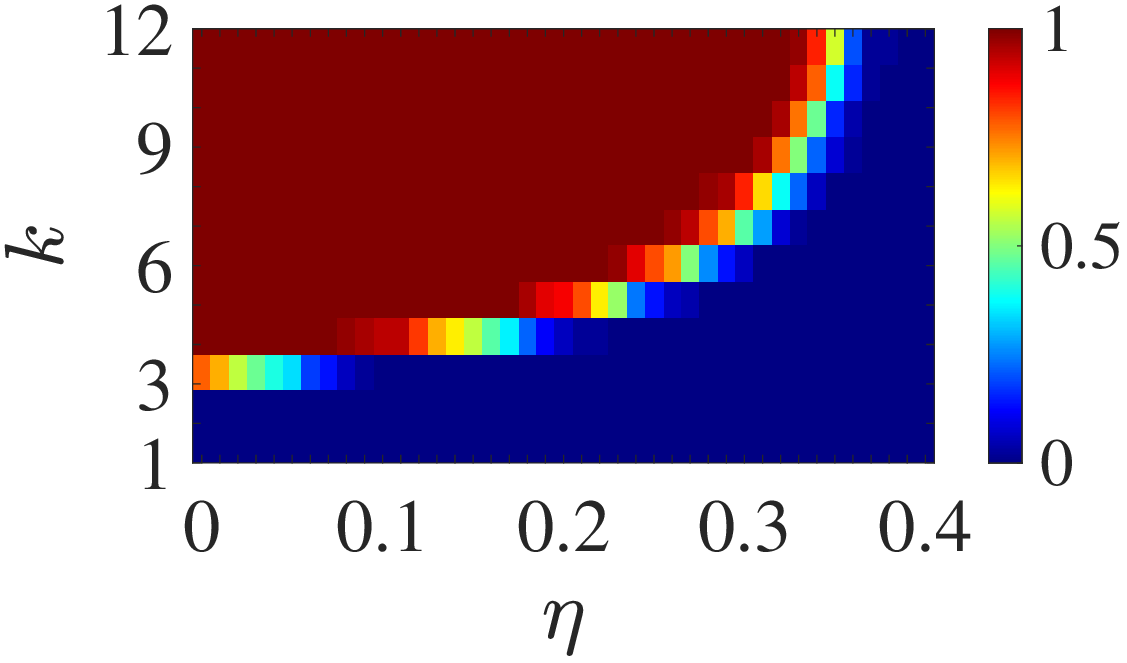}

        \vspace{0.2cm}

        \includegraphics[width=0.45\linewidth]{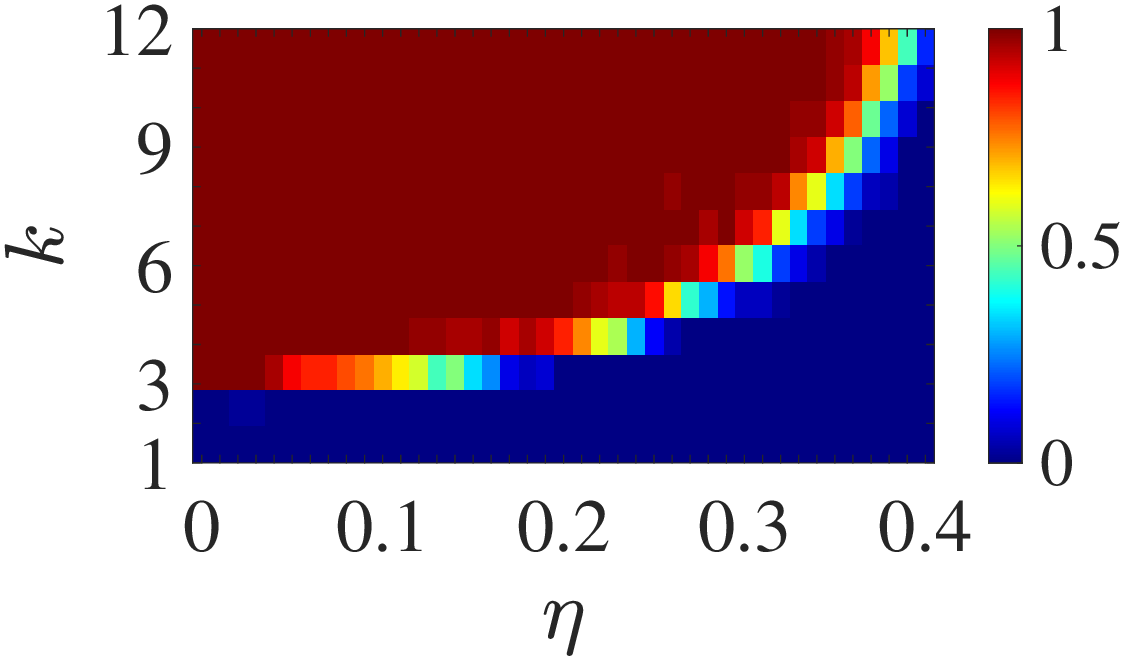}
        \hfill
        \includegraphics[width=0.45\linewidth]{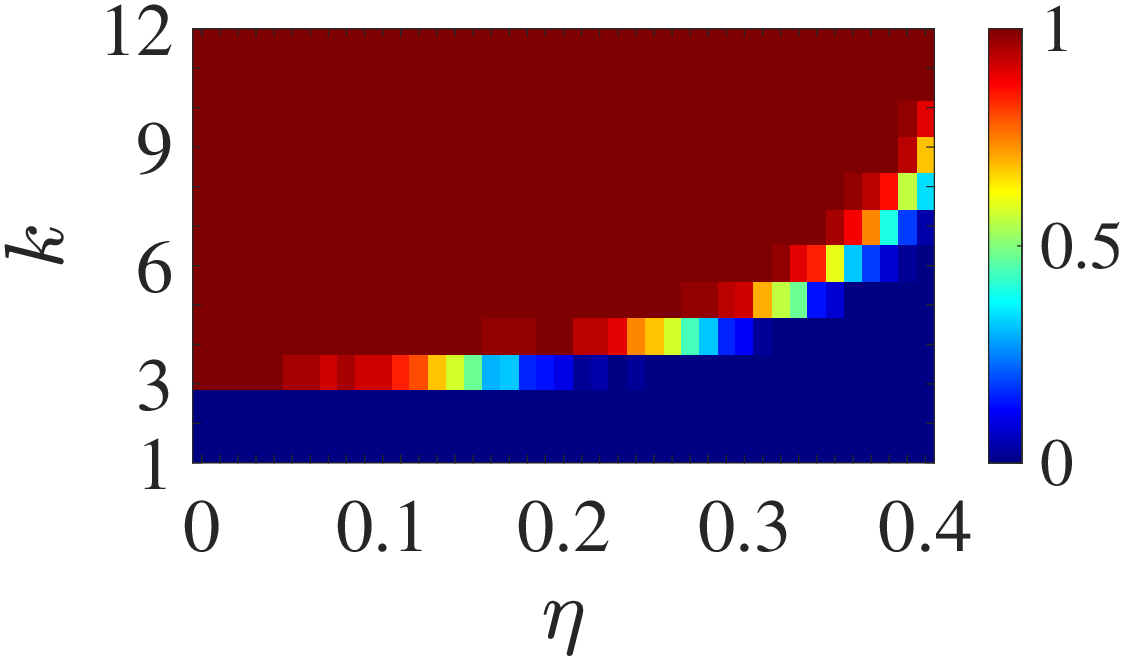}
    \end{subfigure}
    \caption{Phase transition of success rate per $k$ and the fraction of outliers $\eta$ for zero outlier magnitude models. 
    Subfigures are displayed according to RobustPhaseMax (top-left), Median-RWF (top-right), prox-linear method (bottom-left), and Robust-AM (bottom-right).} 
\label{fig:real_phase}
\end{figure}

Robust-AM and the competing algorithms are tested on the collection of $50$ images of handwritten digits\footnote{\url{https://hastie.su.domains/ElemStatLearn/datasets/zip.digits}.}
\Cref{fig:real_phase} compares the two methods in the empirical success rate over $50$ images, where the number of random modulations $k$ and the outlier fraction $\eta$ respectively vary over $k\in\{1,\ldots,12\}$ and $\eta\in[0,0.4]$. 
Similar to the previous experiments on synthetic data, \Cref{fig:real_phase} demonstrates that Robust-AM outperforms the competing algorithms by providing recovery with smaller $k$ for each observed $\eta$.
Since the algorithmic parameters of Median-RWF are specifically selected for Gaussian measurements in \cite{zhang2018median}, we heuristically tuned the step size to $0.2$ so that Median-RWF performs for the measurement setting \eqref{eq:structured_A}.


}

\section{Proof of \Cref{thm:main}}
\label{sec:proof_mainthm}
We first prove by the induction on the iteration index $j$ that 
\begin{equation}
\label{eq:reduction_arg}
    \dist{\mb x_{j},\mb x_\star}\leq\nu_\eta\cdot\dist{\mb x_{j-1},\mb x_\star}+\frac{\epsilon_{j-1}}{C_\eta}
\end{equation} 
holds for all $j \in \mathbb{N}$ for some numerical constant $\nu_\eta \in (0,1)$ and $C_\eta > 0$ depending only on $\eta$. 
Let $k\in\mathbb{N}$ be arbitrarily fixed. Suppose that $\mb x_j$ satisfies \eqref{eq:reduction_arg} for all $j \leq k$. 
Note that the distance between $\mb x$ and $\mb x_\star$ is written as 
\begin{equation}
\label{eq:dist_eq_euclidean}
\mathrm{dist}(\mb x,\mb x_\star)=\|\mb x-\varphi(\mb x)\mb x_\star\|_2,
\end{equation}
where 
\[
\varphi (\mb x) := \argmin_{\alpha\in\{\pm1\}}\euclnorm{\mb x-\alpha\mb x_\star}.
\] 
Then we have $\dist{\mb x_{k+1},\mb x_\star}{\leq}\|\mb x_{k+1}-\varphi(\mb x_k)\mb x_\star\|_2$ and $\mathrm{dist}(\mb x_k,\mb x_\star)=\|\mb x_k-\varphi(\mb x_k)\mb x_\star\|_2$. 
Therefore, it follows that 
\begin{equation}
\label{eq:fstargment}
\|\mb x_{k+1}-\varphi(\mb x_k)\mb x_\star\|_2\leq \nu_\eta\|\mb x_k-\varphi(\mb x_k)\mb x_\star\|_2+\frac{\epsilon_k}{C_\eta}
\end{equation}
implies \eqref{eq:reduction_arg} for $j=k+1$. 
This completes the induction argument. 

Therefore, it suffices to show that the hypothesis of the theorem implies \eqref{eq:fstargment}. 
For the sake of brevity, we denote the objective function of the optimization formulation in \eqref{eq:Robust-AM_explicit} by
\begin{align*}
f_{\mb x_k}(\mb x) & =\frac{1}{m}\sum_{i=1}^{m}\left|\mathrm{sign}\left(\langle\mb a_i,\mb x_k\rangle\right)\langle\mb a_i,\mb x\rangle-b_i\right|.
\end{align*}
Then \eqref{eq:inexact_min} provides
\begin{equation}
\label{eq:optimal_cond}
\underbrace{f_{\mb x_k}(\mb x_{k+1})}_{\mathrm{(A)}}\leq \underbrace{f_{\mb x_k}(\varphi(\mb x_k)\mb x_{\star})}_{\mathrm{(B)}}+\epsilon_k.
\end{equation} 
Next, we derive a lower bound (resp. an upper bound) on (A) (resp. (B)) of \eqref{eq:optimal_cond}.
By from the definition of $b_i$ in \eqref{eq:measurements}, (A) is written as
\begin{equation}
\label{eq:lhs1}
\begin{aligned}
\mathrm{(A)}&=\frac{1}{m}\sum_{i=1}^{m}\left|\mathrm{sign}\left(\langle\mb a_i,\mb x_k\rangle\right)\langle\mb a_i,\mb x_{k+1}\rangle-b_i\right|\\
&=\underbrace{\frac{1}{m}\sum_{i\in I_{\mathrm{in}}}\left|\mathrm{sign}(\langle\mb a_i,\mb x_k\rangle)\langle\mb a_i,\mb x_{k+1}\rangle-|\langle\mb a_i,\varphi(\mb x_k)\mb x_\star\rangle|\right|}_{\mathrm{(a)}}\\
&\quad+\frac{1}{m}\sum_{i\in I_{\mathrm{out}}}\left|\mathrm{sign}\left(\langle\mb a_i,\mb x_k\rangle\right)\langle\mb a_i,\mb x_{k+1}\rangle-\xi_i\right|.
\end{aligned}
\end{equation} 
To simplify the partial summation over $I_{\mathrm{in}}$, we introduce the spherical wedge \cite{tan2019phase} defined by
\begin{equation}
\label{eq:def_wedge}
W_{\mb x,\mb z}:=\{\mb v\in\mathbb{S}^{d-1} \,|\, \mathrm{sign}(\langle\mb v,\mb x\rangle)\neq\mathrm{sign}(\langle\mb v,\mb z\rangle)\}.
\end{equation} 
Then if follows that $\langle\mb a_i,\varphi(\mb x_k)\mb x_\star\rangle$ and $\langle\mb a_i,\mb x_k\rangle$ have the opposite sign if and only if $\bm a_i \in W_{\mb x_k,\varphi(\mb x_k)\mb x_\star}$. 
Therefore, the summand in (a) is rewritten as 
\begin{align*}
\mathrm{(a)} 
&=\frac{1}{m}\sum_{i\in I_{\mathrm{in}}}\bbone_{\{\mb a_i\in W_{\mb x_k,\varphi(\mb x_k)\mb x_\star}\}}\left|\langle\mb a_i,\mb x_{k+1}+\varphi(\mb x_k)\mb x_\star\rangle\right|\\
&\quad+\frac{1}{m}\sum_{i\in I_{\mathrm{in}}}\bbone_{\{\mb a_i\notin W_{\mb x_k,\varphi(\mb x_k)\mb x_\star}\}}\left|\langle\mb a_i,\mb x_{k+1}-\varphi(\mb x_k)\mb x_\star\rangle\right|. 
\end{align*}
The second summand on the right-hand side provides a valid lower bound on (a) since the other summand is nonnegative. 
Combining the above results, we obtain that 
\begin{equation}
\label{eq:lhs1}
\begin{aligned}
\mathrm{(A)}&\geq\frac{1}{m}\sum_{i\in I_{\mathrm{in}}}\bbone_{\{\mb a_i\notin W_{\mb x_k,\varphi(\mb x_k)\mb x_\star}\}}\left|\langle\mb a_i,\mb x_{k+1}-\varphi(\mb x_k)\mb x_\star\rangle\right|\\
&\quad+\frac{1}{m}\sum_{i\in I_{\mathrm{out}}}\left|\mathrm{sign}\left(\langle\mb a_i,\mb x_k\rangle\right)\langle\mb a_i,\mb x_{k+1}\rangle-\xi_i\right|.
\end{aligned}
\end{equation} 
Similarly, (B) is written as 
\begin{align*}
\mathrm{(B)}&{=}\frac{1}{m}\sum_{i\in I_{\mathrm{in}}} \underbrace{\left|\mathrm{sign}(\langle\mb a_i,\mb x_k\rangle)\langle\mb a_i,\varphi(\mb x_k)\mb x_\star\rangle-|\langle\mb a_i,\varphi(\mb x_k)\mb x_\star\rangle|\right|}_{\mathrm{(b)}}\\
&\quad+\frac{1}{m}\sum_{i\in I_{\mathrm{out}}}\left|\mathrm{sign}(\langle\mb a_i,\mb x_k\rangle)\langle\mb a_i,\varphi(\mb x_k)\mb x_\star\rangle-\xi_i\right|.
\end{align*}
If $\bm a_i \in W_{\mb x_k,\varphi(\mb x_k)\mb x_\star}$, then $\langle\mb a_i,\mb x_k\rangle$ and $\langle\mb a_i,\varphi(\mb x_k)\mb x_\star\rangle$ have the opposite sign and hence (b) satisfies
\begin{align*}
\mathrm{(b)} 
&= 2 \left|\langle\mb a_i,\mb x_\star\rangle\right| 
\leq 2\left|\langle\mb a_i,\varphi(\mb x_k)\mb x_\star-\mb x_k\rangle\right|.
\end{align*}
Otherwise, if $\bm a_i \not\in W_{\mb x_k,\varphi(\mb x_k)\mb x_\star}$, then $\mathrm{(b)} = 0$. 
Therefore, we have
\begin{equation}
\label{eq:rhs1}
\begin{aligned}
\mathrm{(B)}&\leq\frac{2}{m}\sum_{i\in I_{\mathrm{in}}}\bbone_{\{\mb a_i\in W_{\mb x_k,\varphi(\mb x_k)\mb x_\star}\}}\left|\langle\mb a_i,\varphi(\mb x_k)\mb x_\star-\mb x_k\rangle\right|\\
&\quad+\frac{1}{m}\sum_{i\in I_{\mathrm{out}}}\left|\mathrm{sign}(\langle\mb a_i,\mb x_k\rangle)\langle\mb a_i,\varphi(\mb x_k)\mb x_\star\rangle-\xi_i\right|.
\end{aligned}
\end{equation} 
By plugging in the bounds of \eqref{eq:lhs1} and \eqref{eq:rhs1} into \eqref{eq:optimal_cond}, we obtain that \eqref{eq:optimal_cond} implies
\begin{equation} 
\label{eq:conclude_ineq1}
\begin{aligned}
&\frac{1}{m}\sum_{i\in I_{\mathrm{in}}}\bbone_{\{\mb a_i\notin W_{\mb x_k,\varphi(\mb x_k)\mb x_\star}\}}\left|\langle\mb a_i,\mb x_{k+1}-\varphi(\mb x_k)\mb x_\star\rangle\right|\\
&\qquad+\underbrace{\frac{1}{m}\sum_{i\in I_{\mathrm{out}}}\left|\mathrm{sign}\left(\langle\mb a_i,\mb x_k\rangle\right)\langle\mb a_i,\mb x_{k+1}\rangle-\xi_i\right|}_{(*)}\\
&\quad\quad-\underbrace{\frac{1}{m}\sum_{i\in I_{\mathrm{out}}}\left|\mathrm{sign}(\langle\mb a_i,\mb x_k\rangle)\langle\mb a_i,\varphi(\mb x_k)\mb x_\star\rangle-\xi_i\right|}_{(**)}\\
&\leq\frac{2}{m}\sum_{i\in I_{\mathrm{in}}}\bbone_{\{\mb a_i\in W_{\mb x_k,\varphi(\mb x_k)\mb x_\star}\}}\left|\langle\mb a_i,\varphi(\mb x_k)\mb x_\star-\mb x_k\rangle\right|+\epsilon_k.
\end{aligned}
\end{equation}
By applying the triangle inequality to the summands in $(*)$ and $(**)$, we obtain a necessary condition of \eqref{eq:conclude_ineq1} given by
\begin{equation}
\label{eq:last_condition}
\begin{aligned}
&\underbrace{\frac{1}{m}\sum_{i\in I_{\mathrm{in}}}\bbone_{\{\mb a_i\notin W_{\mb x_k,\varphi(\mb x_k)\mb x_\star}\}}\left|\langle\mb a_i,\mb x_{k+1}-\varphi(\mb x_k)\mb x_\star\rangle\right|}_{\mathrm{(c)}}\\
&\qquad\qquad\qquad\qquad-\underbrace{{\frac{1}{m}\sum_{i\in I_{\mathrm{out}}}\left|\langle\mb a_i,\mb x_{k+1}-\varphi(\mb x_k)\mb x_\star\rangle\right|}}_{\mathrm{(d)}}\\
&\leq\underbrace{\frac{2}{m}\sum_{i\in I_{\mathrm{in}}}\bbone_{\{\mb a_i\in W_{\mb x_k,\varphi(\mb x_k)\mb x_\star}\}}\left|\langle\mb a_i,\varphi(\mb x_k)\mb x_\star-\mb x_k\rangle\right|}_{\mathrm{(e)}}+\epsilon_k.
\end{aligned}
\end{equation}

We have shown that \eqref{eq:optimal_cond} implies \eqref{eq:last_condition}. 
In the remainder of the proof, we demonstrate that if \eqref{eq:last_condition} is satisfied, then \eqref{eq:fstargment} holds with high probability. This is achieved by applying a probabilistic lower bound on (c) and probabilistic upper bounds on (d) and (e), using 
concentration inequalities.

To this end, note that the measurement vectors $\{\mb a_i\}_{i=1}^m$ depend not only on the current iterate $\mb x_k$ and the next iterate $\mb x_{k+1}$, but also on the indication functions within the spherical wedge in (c) and (e). Therefore, we consider the uniform bounds for all iterates and the collection of spherical wedges with the largest angle less than $\theta \in (0, \pi)$. We introduce the corresponding lemmas below.

\begin{lemma}
\label{lem:keylemma}
Let  $\theta\in(0,\pi), \eta\in(0,1/2)$ and $\delta>0$. Suppose that $\{\mb a_i\}_{i=1}^{m}$ are independent copies of $\mb g \sim \mathrm{Normal}(\bm 0, \bm I_d)$. Let 
\begin{equation}
\label{eq:def_Wtheta}
\mathcal{W}_{\theta}:=\left\{W_{\mb x,\mb z} : \mb x,\mb z\in\mathbb{R}^d, \angle\left(\mb x,\mb z\right)\leq\theta \right\},
\end{equation} where $W_{\mb x,\mb z}$ is defined in \eqref{eq:def_wedge}.
Then there exists an absolute constant $C$ such that 
\begin{align}
&\inf_{\begin{subarray}{l} W \in \mathcal{W}_{\theta} \\ \mb z\in\mathbb{S}^{d-1}\end{subarray}}{\frac{1}{m}\sum_{i\in I_{\mathrm{in}}}\bbone_{\{\mb a_i\notin W\}}\left|\langle\mb a_i,\mb z\rangle\right|}\geq (1-\eta)\sqrt{\frac{2}{\pi}}\nonumber\\
&-\frac{2\theta}{\pi}\left(\sqrt{\frac{2}{\pi}}+\sqrt{2\log\left(\frac{e\pi}{2\theta}\right)}\right)-\frac{\theta}{20}\left(\sqrt{\frac{2\theta}{\pi}}+1\right),\label{eq:key_lowerbound}\\
&\sup_{\mb z\in\mathbb{S}^{d-1}}\frac{1}{m}\sum_{i\in I_{\mathrm{out}}}|\langle\mb a_i,\mb z\rangle|\leq\eta\sqrt{\frac{\pi}{2}}+\sqrt{\eta}\frac{\theta}{20},
\label{eq:key_lowerbound2}\\
&\mathrm{and}\nonumber\\
&\sup_{\begin{subarray}{l} W \in \mathcal{W}_{\theta} \\ \mb z\in\mathbb{S}^{d-1}\end{subarray}}\frac{1}{m}\sum_{i\in I_{\mathrm{in}}}\bbone_{\{\mb a_i\in W\}}\left|\langle\mb a_i,\mb z\rangle\right|\nonumber\\
&\leq \frac{2\theta}{\pi}\left(\sqrt{\frac{2}{\pi}}+\sqrt{2\log\left(\frac{e\pi}{2\theta}\right)}\right)+\sqrt{\frac{2\theta}{\pi}}\cdot\frac{\theta}{20}
\label{eq:key_lowerbound3}
\end{align} hold with probability at least $1-\delta$ provided that
\begin{equation}
\label{eq:sample_main_lemma}
m\geq C\cdot\theta^{-2}\left(d\log(m/d)\vee\log(1/\delta)\right).
\end{equation}
\end{lemma}
\begin{IEEEproof}
See \Cref{sec:proof_mainLemma}.
\end{IEEEproof}
Now we derive the largest angle for the spherical wedge $W_{\mb x_k,\varphi(\mb x_k)\mb x_\star}$. Since the angle between $\mb x_k$ and $\varphi(\mb x_k)\mb x_\star$ is always acute, we have
\begin{equation}
\label{eq:distance_upperbound}
\begin{aligned}
& \sin\left(\angle\left(\mb x_k,\varphi(\mb x_k)\mb x_\star\right)\right)
=\left\|\left(\mb I_{d}-\frac{\mb x_k\mb x_k^\T}{\|\mb x_k\|_2^2}\right)\frac{\varphi(\mb x_k)\mb x_\star}{\|\mb x_\star\|_2}\right\|\\ 
&\quad\leq\left\|\left(\mb I_{d}-\frac{\mb x_k\mb x_k^\T}{\|\mb x_k\|_2^2}\right)\frac{\varphi(\mb x_k)\mb x_\star-\mb x_k}{\|\mb x_\star\|_2}\right\|\\
&\quad \overset{\mathrm{(i)}}{\leq}\frac{\|\mb x_k-\varphi(\mb x_k)\mb x_\star\|_2}{\|\mb x_\star\|_2}
= \frac{\dist{\mb x_{k},\mb x_\star}}{\|\mb x_\star\|_2} \\
&\quad \overset{\mathrm{(ii)}}{\leq}\sin\left(\frac{2}{25}\right),
\end{aligned} 
\end{equation}
where (i) holds since the project operator is non-expansive; (ii) follows since the induction hypothesis implies
{
\[
\begin{aligned}
&\dist{\mb x_k,\mb x_\star}\\
&\leq \nu_{\eta}^k \cdot \dist{\mb x_0,\mb x_\star}+\frac{\max_{i\in[0:k-1]}\epsilon_{i}}{C_\eta}\sum_{t=0}^{k-1}\nu_\eta^t\\
&\leq\nu_{\eta}^k \cdot \dist{\mb x_0,\mb x_\star}+(1-\nu_\eta)\sin\left(\frac{2}{25}\right)\|\mb x_\star\|_2\sum_{t=0}^{k-1}\nu_\eta^t\\
&\leq\sin\left(\frac{2}{25}\right)\|\mb x_\star\|_2,
\end{aligned}
\] where the second and the last inequalities follow from \eqref{eq:initial_condition}.
}

Hence, in \Cref{lem:keylemma}, we plug in $\theta=2/25$. Then the sample complexity in \Cref{thm:main} invokes \Cref{lem:keylemma}, \eqref{eq:key_lowerbound}, \eqref{eq:key_lowerbound2}, and \eqref{eq:key_lowerbound3} hold with probability at least $1-\delta$ simultaneously. The remainder of the proof is conditioned on the events that \eqref{eq:key_lowerbound}, \eqref{eq:key_lowerbound2}, and \eqref{eq:key_lowerbound3} hold.

By applying \eqref{eq:key_lowerbound} and \eqref{eq:key_lowerbound2} to (c) and (d) of \eqref{eq:last_condition} and \eqref{eq:key_lowerbound3} to (e) of \eqref{eq:last_condition} with the choice of $\theta=2/25$, we obtain
\[
\|\mb x_{k+1}-\varphi(\mb x_k)\mb x_\star\|_2 
\leq 
\nu_\eta \|\mb x_k-\varphi(\mb x_k)\mb x_\star\|_2 +\frac{\epsilon_k}{C_\eta}
\]
for 
\begin{equation}
\label{eq:def_nueta}
\nu_\eta := \frac{c_0}{C_\eta} \quad\text{and}\quad
C_\eta:=(1-2\eta)\sqrt{\frac{2}{\pi}}-c_0-\frac{1}{250}(1+\sqrt{\eta}), 
\end{equation}
where 
\[
c_0:=\frac{4}{25\pi}\left(\sqrt{\frac{2}{\pi}}+\sqrt{2\log\left(\frac{25e\pi}{4}\right)}\right)+\frac{1}{625\sqrt{\pi}}.
\] 
Since $\nu_\eta$ satisfies
\[
\frac{d\nu_\eta}{d\eta}=\frac{c_0\left(2\sqrt{\frac{2}{\pi}}+\frac{1}{500\sqrt{\eta}}\right)}{\left((1-2\eta)\sqrt{\frac{2}{\pi}}-c_0-\frac{1}{250}(1+\sqrt{\eta})\right)^2}>0
\]
for all $\eta\in[0,1/4]$, it is monotonically increasing in $\eta$ and upper-bounded as $\nu_\eta \leq \nu_{1/4}<9/10$. This implies $\nu_\eta<1$ uniformly over $\eta\in[0,1/4]$. This completes the proof of \eqref{eq:fstargment}.





\section{Supporting Lemmas}

\begin{lemma}
\label{lem:bounds_probs}
Let $\mb g\sim\mathrm{Normal}(\mb 0,\mb I_d)$ and $\theta\in(0,\pi)$. Let $\mathcal{W}_\theta$ be defined as in \eqref{eq:def_Wtheta}. Then we have
\[
\sup_{W\in\mathcal{W}_\theta}\P(\mb g\in W)\leq\frac{\theta}{\pi}.
\]
\end{lemma}
\begin{IEEEproof}
Let $W\in\mathcal{W}_{\theta}$ be arbitrarily fixed. 
It follows from the definitions in \eqref{eq:def_Wtheta} and \eqref{eq:def_wedge} that $W$ is a cone. 
Therefore, $\mb g \in W$ if and only if $\mb g/\|\mb g\|_2 \in W$. 
Furthermore, note that $\mb g/\|\mb g\|_2$ is uniformly distributed in $\mathbb{S}^{d-1}$. 
Then we have 
\begin{equation}
\label{eq:lowerbnd_prob_theta}
\P\left(\mb g\in W\right)=\P\left(\frac{\mb g}{\|\mb g\|_2} \in W\right)\leq\frac{\theta}{\pi}. 
\end{equation}
The assertion follows since $W$ was arbitrary. 
\end{IEEEproof}

\begin{lemma}[{\hspace{1sp}\cite[Lemma~2.1]{plan2014dimension}}]
\label{lem:bounds_probs} 
Let $\delta\in(0,1)$ and $\{\mb a_i\}_{i=1}^{m}$ be independent copies of $\mb g \sim \mathrm{Normal}(\mb 0, \mb I_d)$. Then it holds with probability at least $1-\delta$ that
\begin{equation}
\label{eq:sup_ell1}
\sup_{\mb z\in S^{d-1}}\left|\frac{1}{m}\sum_{i=1}^{m}|\langle\mb a_i,\mb z\rangle|-\sqrt{\frac{2}{\pi}}\right|\leq4\sqrt{\frac{d}{m}}+\sqrt{\frac{2\log(2/\delta)}{m}}.
\end{equation}
\end{lemma}

\begin{lemma}[{\hspace{1sp}\cite[Lemma~6.4]{plan2012robust}}]
\label{lem:bounds_probs_partial} 
Let $\delta\in(0,1)$ and $\{\mb a_i\}_{i=1}^{m}$ be independent copies of $\mb g \sim \mathrm{Normal}(\mb 0, \mb I_d)$. 
Let $s \in \mathbb{N}$ satisfy $s<m$. Then it holds with probability at least $1-\delta$ that
\begin{equation}
\label{eq:sup_ell1_partial}
\begin{aligned}
& \sup_{\begin{subarray}{l} \mb z \in \mathbb{S}^{d-1} \\ T:|T|\leq s \end{subarray}} \frac{1}{s} \sum_{i\in T}|\langle\mb a_i,\mb z\rangle| \\
& \quad \leq \sqrt{\frac{2}{\pi}}
+4\sqrt{\frac{d}{s}}+\sqrt{2\log\left(\frac{em}{s}\right)}+\sqrt{\frac{2}{s} \cdot \log\left(\frac{2}{\delta}\right)}.
\end{aligned}
\end{equation} 
\end{lemma}

\begin{lemma}[{\hspace{1sp}\cite[Lemma~5.1]{tan2019phase}}]
\label{lem:emprical_upbnd1}
Let $\delta\in(0,1)$ and an acute angle $\theta>0$. Suppose $\{\mb a_i\}_{i=1}^{m}$ be independent copies of a random variable $\mb a\in \mathbb{R}^d$ and we consider the set $\mathcal{W}_\theta$ given by \eqref{eq:def_Wtheta}. Then, if
\[
m\geq (4\pi/\theta)^2(2d\log(2em/d)+\log(2/\delta)),
\] we have
\begin{equation}
\sup_{W\in\mathcal{W}_\theta}\frac{1}{m}\sum_{i=1}^{m}\bbone_{\{\mb a_i\in W\}}\leq \frac{2\theta}{\pi}.
\end{equation} holds with probability at least $1-\delta$.
\end{lemma}

\section{Proof of \Cref{lem:keylemma}}
\label{sec:proof_mainLemma}
We proceed with the proof under the following four events, each of which holds with probability at least $1-\delta/4$. The first event is defined as
\begin{equation}
\label{eq:first_event}
\begin{aligned}
&\sup_{\mb z\in\mathbb{S}^{d-1}}\left|\frac{1}{m}\sum_{i\in I_{\mathrm{in}}}|\langle\mb a_i,\mb z\rangle|-(1-\eta)\sqrt{\frac{2}{\pi}}\right|\\
&\qquad\qquad\leq4\sqrt{\frac{d}{m}}+\sqrt{\frac{2\log(8/\delta)}{m}},\\
\end{aligned}
\end{equation} 
which holds with probability at least $1-\delta/4$.
Since by the assumption on outliers, we have a set $|I_{\mathrm{in}}|$ with $|I_{\mathrm{in}}|=(1-\eta)m$ and the outliers are independent of $\{\mb a_i\}_{i=1}^{m}$. Hence, \eqref{eq:first_event} is a direct result of \eqref{eq:sup_ell1} in \Cref{lem:bounds_probs}. By following the same argument, we also have that
\begin{equation}
\label{eq:secondevent}
\sup_{\mb z\in\mathbb{S}^{d-1}}\left|\frac{1}{m}\sum_{i\in I_{\mathrm{out}}}|\langle\mb a_i,\mb z\rangle|-\eta\sqrt{\frac{2}{\pi}}\right|\leq4\sqrt{\frac{\eta d}{m}}+\sqrt{\frac{2 \eta \log(8/\delta)}{m}}
\end{equation} 
holds with probability at least $1-\delta/4$.

Next, we describe the following event: for an arbitrary fixed $\alpha\in(0,1)$, it holds with probability at least $1-\delta/4$ that
\begin{equation}
\label{eq:thirdevent}
\begin{aligned}
&\sup_{\begin{subarray}{l}
T:|T|\leq \alpha m \\ \mb z\in\mathbb{S}^{d-1}    
\end{subarray}}\frac{1}{m}\sum_{i\in T \cap I_{\mathrm{in}}} \left|\langle\mb a_i,\mb z\rangle\right|\leq\\
&\alpha\sqrt{\frac{2}{\pi}}+4\sqrt{\frac{\alpha d}{m}}+\alpha\sqrt{2\log\left(\frac{e}{\alpha}\right)}+\sqrt{\frac{2\alpha\log(8/\delta)}{m}}.
\end{aligned}
\end{equation} Again, since by the Assumption~1, we have a fixed set $|I_{\mathrm{in}}|$ with $|I_{\mathrm{in}}|=(1-\eta)m$ and the outliers are independent of $\{\mb a_i\}_{i=1}^{m}$, 
\eqref{eq:thirdevent} holds by \eqref{eq:sup_ell1_partial} in \Cref{lem:bounds_probs_partial}.
 
Since \eqref{eq:sample_main_lemma} invokes \Cref{lem:emprical_upbnd1} with probability at least $1-\delta/4$, it holds with probability at least $1-\delta/4$ that
\begin{equation}
\label{eq:fourthevent}
\sup_{W\in\mathcal{W}_\theta}\sum_{i=1}^{m}\bbone_{\{\mb a_i\in W\}}\leq\frac{2\theta m}{\pi}.
\end{equation} 

Since we have shown that \eqref{eq:first_event},\eqref{eq:secondevent},\eqref{eq:thirdevent} and \eqref{eq:fourthevent} hold with probability
at least $1-\delta$, we will move forward with the remainder of the proof by assuming those conditions are satisfied.

We first show \eqref{eq:key_lowerbound}. We observe that for an arbitrary $W\in\mathcal{W}_\theta$ and $\mb z\in\mathbb{S}^{d-1}$, it holds deterministically that 
\begin{equation*}
\begin{aligned}
\frac{1}{m}&\sum_{i\in I_{\mathrm{in}}}\bbone_{\{\mb a_i\notin W\}}|\langle\mb a_i,\mb z\rangle|=\\
&\frac{1}{m}\sum_{i\in I_{\mathrm{in}}}|\langle\mb a_i,\mb z\rangle|-\frac{1}{m}\sum_{i\in I_{\mathrm{in}}}\bbone_{\{\mb a_i \in W\}}|\langle\mb a_i,\mb z\rangle|.
\end{aligned}
\end{equation*}
Hence, by taking infimum on both sides over sets $\mathcal{W}_\theta$ and $\mathbb{S}^{d-1}$, we have
\begin{equation}
\label{eq:decompose_1_notin}
\begin{aligned}
&\inf_{\begin{subarray}{l}
    W\in\mathcal{W}_\theta \\ \mb z\in\mathbb{S}^{d-1}
\end{subarray}}\frac{1}{m}\sum_{i\in I_{\mathrm{in}}}\bbone_{\{\mb a_i\notin W\}}|\langle\mb a_i,\mb z\rangle| \\
& \geq\underbrace{\inf_{\mb z\in\mathbb{S}^{d-1}}\frac{1}{m}\sum_{i\in I_{\mathrm{in}}}|\langle\mb a_i,\mb z\rangle|}_{\mathrm{(A)}}
-\underbrace{\sup_{\begin{subarray}{l}
    W\in\mathcal{W}_\theta \\ \mb z\in\mathbb{S}^{d-1}
\end{subarray}}\frac{1}{m}\sum_{i\in I_{\mathrm{in}}}\bbone_{\{\mb a_i\in W\}}|\langle\mb a_i,\mb z\rangle|}_{\mathrm{(B)}}.
\end{aligned}
\end{equation}
We first obtain a lower bound on (A) and an upper bound on (B). We have a lower bound on (A) by \eqref{eq:first_event}:
\begin{equation}
\label{eq:Alwbbnd}
\mathrm{(A)}\geq (1-\eta)\sqrt{\frac{2}{\pi}}-4\sqrt{\frac{d}{m}} - \sqrt{\frac{2\log(8/\delta)}{m}}.
\end{equation}
By taking $m$ \eqref{eq:sample_main_lemma} in \eqref{eq:Alwbbnd} for a sufficiently large $C>0$, we have 
\begin{equation}
\label{eq:resultA}
\mathrm{(A)}\geq(1-\eta)\sqrt{\frac{2}{\pi}}-\frac{\theta}{20}.
\end{equation} 
It remains to show an upper bound on (B). Under the event \eqref{eq:fourthevent}, we have
\[
\mathrm{(B)}\leq\sup_{\begin{subarray}{l}
T:|T|\leq{2\theta m}/\pi \\ \mb z\in\mathbb{S}^{d-1}    
\end{subarray}}\frac{1}{m}\sum_{i\in T \cap I_{\mathrm{in}}} \left|\langle\mb a_i,\mb z\rangle\right|.
\]
Therefore, by letting $\alpha=2\theta/\pi$ in \eqref{eq:thirdevent}, \eqref{eq:thirdevent} gives an upper bound on (B):
\begin{equation}
\label{eq:Bupbnd}
\begin{aligned}
&\mathrm{(B)}\leq\frac{2\theta}{\pi}\sqrt{\frac{2}{\pi}}+4\sqrt{\frac{2\theta d}{\pi m}}
+\frac{2\theta}{\pi}\sqrt{2\log\left(\frac{e\pi}{2\theta}\right)}+\sqrt{\frac{4\theta \log(8/\delta)}{\pi m}}.
\end{aligned}
\end{equation}
Taking $m$ according to \eqref{eq:sample_main_lemma} yields 
\begin{equation}
\label{eq:resultB}
\mathrm{(B)}\leq\frac{2\theta}{\pi}\left(\sqrt{\frac{2}{\pi}}+\sqrt{2\log\left(\frac{e\pi}{2\theta}\right)}\right)+\frac{\theta}{20}\sqrt{\frac{2\theta}{\pi}}.
\end{equation} 

Hence, putting the results \eqref{eq:resultA} and \eqref{eq:resultB} into \eqref{eq:decompose_1_notin} completes the proof of the statement \eqref{eq:key_lowerbound}. 

For the proofs of remaining statements in \eqref{eq:key_lowerbound2} and \eqref{eq:key_lowerbound3}, the upper bound in \eqref{eq:key_lowerbound2} is a direct consequence of \eqref{eq:secondevent} with choosing $n$ according   \eqref{eq:sample_main_lemma}. Lastly, \eqref{eq:key_lowerbound3} is the result of the upper bound of (B) in \eqref{eq:resultB}. These complete the proof of \eqref{eq:key_lowerbound2} and \eqref{eq:key_lowerbound3}.

\section{Conclusion}
\label{sec:discussion}

The least absolute deviation (LAD) has been a popular statistical method for regression in the presence of outliers. We consider the LAD approach to robust phase retrieval with the magnitude-only measurement model. To solve the resulting non-convex optimization, we derive a robust alternating minimization method (Robust-AM) as an unconstrained Gauss-Newton method. 
Furthermore, we propose fast Robust-AM by exploiting efficient solvers and show that Robust-AM by ADMM converges faster than a similar approach known as the prox-linear by its efficient solver POGS \cite{duchi2019solving}.

We established a local convergence analysis of Robust-AM under the standard Gaussian measurement model when the support of sparse noise is arbitrarily fixed but magnitudes can be adversarial. 
A suitably initialized Robust-AM converges linearly to the ground truth uniformly over all ground-truth signals when the number of measurements $m$ is proportional to the signal length $d$ and the outlier fraction is up to $1/4$. 
This theoretical result is comparable to existing prior art in the literature. 
Furthermore, the numerical results show that Robust-AM outperforms the existing guaranteed methods for various outlier models in both synthetic data and real image data.

\bibliographystyle{IEEEtran}
\bibliography{ref}

\begin{thebibliography}{10}
\providecommand{\url}[1]{#1}
\csname url@samestyle\endcsname
\providecommand{\newblock}{\relax}
\providecommand{\bibinfo}[2]{#2}
\providecommand{\BIBentrySTDinterwordspacing}{\spaceskip=0pt\relax}
\providecommand{\BIBentryALTinterwordstretchfactor}{4}
\providecommand{\BIBentryALTinterwordspacing}{\spaceskip=\fontdimen2\font plus
\BIBentryALTinterwordstretchfactor\fontdimen3\font minus \fontdimen4\font\relax}
\providecommand{\BIBforeignlanguage}[2]{{%
\expandafter\ifx\csname l@#1\endcsname\relax
\typeout{** WARNING: IEEEtran.bst: No hyphenation pattern has been}%
\typeout{** loaded for the language `#1'. Using the pattern for}%
\typeout{** the default language instead.}%
\else
\language=\csname l@#1\endcsname
\fi
#2}}
\providecommand{\BIBdecl}{\relax}
\BIBdecl

\bibitem{kim2024robust}
S.~Kim and K.~Lee, ``Sequence of linear program for robust phase retrieval,'' \emph{2024 IEEE International Conference on Acoustics, Speech and Signal Processing}, to appear.

\bibitem{walther1963question}
A.~Walther, ``The question of phase retrieval in optics,'' \emph{Optica Acta: International Journal of Optics}, vol.~10, no.~1, pp. 41--49, 1963.

\bibitem{bunk2007diffractive}
O.~Bunk, A.~Diaz, F.~Pfeiffer, C.~David, B.~Schmitt, D.~K. Satapathy, and J.~F. Van Der~Veen, ``Diffractive imaging for periodic samples: retrieving one-dimensional concentration profiles across microfluidic channels,'' \emph{Acta Crystallographica Section A: Foundations of Crystallography}, vol.~63, no.~4, pp. 306--314, 2007.

\bibitem{chai2010array}
A.~Chai, M.~Moscoso, and G.~Papanicolaou, ``Array imaging using intensity-only measurements,'' \emph{Inverse Problems}, vol.~27, no.~1, p. 015005, 2010.

\bibitem{shechtman2015phase}
Y.~Shechtman, Y.~C. Eldar, O.~Cohen, H.~N. Chapman, J.~Miao, and M.~Segev, ``Phase retrieval with application to optical imaging: a contemporary overview,'' \emph{IEEE Signal Processing Magazine}, vol.~32, no.~3, pp. 87--109, 2015.

\bibitem{weller2015undersampled}
D.~S. Weller, A.~Pnueli, G.~Divon, O.~Radzyner, Y.~C. Eldar, and J.~A. Fessler, ``Undersampled phase retrieval with outliers,'' \emph{IEEE Transactions on Computational Imaging}, vol.~1, no.~4, pp. 247--258, 2015.

\bibitem{dong2023phase}
J.~Dong, L.~Valzania, A.~Maillard, T.-a. Pham, S.~Gigan, and M.~Unser, ``Phase retrieval: From computational imaging to machine learning: A tutorial,'' \emph{IEEE Signal Processing Magazine}, vol.~40, no.~1, pp. 45--57, 2023.

\bibitem{bahmani2017phase}
S.~Bahmani and J.~Romberg, ``Phase retrieval meets statistical learning theory: A flexible convex relaxation,'' in \emph{Artificial Intelligence and Statistics}.\hskip 1em plus 0.5em minus 0.4em\relax PMLR, 2017, pp. 252--260.

\bibitem{goldstein2018phasemax}
T.~Goldstein and C.~Studer, ``Phasemax: Convex phase retrieval via basis pursuit,'' \emph{IEEE Transactions on Information Theory}, vol.~64, no.~4, pp. 2675--2689, 2018.

\bibitem{hand2016corruption}
P.~Hand and V.~Voroninski, ``Corruption robust phase retrieval via linear programming,'' \emph{arXiv preprint arXiv:1612.03547}, 2016.

\bibitem{zhang2017nonconvex}
H.~Zhang, Y.~Zhou, Y.~Liang, and Y.~Chi, ``A nonconvex approach for phase retrieval: Reshaped wirtinger flow and incremental algorithms,'' \emph{Journal of Machine Learning Research}, 2017.

\bibitem{wang2017solving}
G.~Wang, G.~B. Giannakis, and Y.~C. Eldar, ``Solving systems of random quadratic equations via truncated amplitude flow,'' \emph{IEEE Transactions on Information Theory}, vol.~64, no.~2, pp. 773--794, 2017.

\bibitem{zhang2018median}
H.~Zhang, Y.~Chi, and Y.~Liang, ``Median-truncated nonconvex approach for phase retrieval with outliers,'' \emph{IEEE Transactions on Information Theory}, vol.~64, no.~11, pp. 7287--7310, 2018.

\bibitem{duchi2019solving}
J.~C. Duchi and F.~Ruan, ``Solving (most) of a set of quadratic equalities: Composite optimization for robust phase retrieval,'' \emph{Information and Inference: A Journal of the IMA}, vol.~8, no.~3, pp. 471--529, 2019.

\bibitem{bloomfield1983least}
P.~Bloomfield and W.~L. Steiger, \emph{Least absolute deviations: theory, applications, and algorithms}.\hskip 1em plus 0.5em minus 0.4em\relax Springer, 1983.

\bibitem{GerchbergS72}
R.~W. Gerchberg and W.~O. Saxton, ``A practical algorithm for the determination of phase from image and diffraction plane pictures,'' \emph{Optik}, vol.~35, p. 237, 1972.

\bibitem{boyd2011distributed}
S.~Boyd, N.~Parikh, E.~Chu, B.~Peleato, and J.~Eckstein, ``Distributed optimization and statistical learning via the alternating direction method of multipliers,'' \emph{Foundations and Trends{\textregistered} in Machine learning}, vol.~3, no.~1, pp. 1--122, 2011.

\bibitem{wang2017new}
S.~Wang and N.~Shroff, ``A new alternating direction method for linear programming,'' \emph{Advances in Neural Information Processing Systems}, vol.~30, 2017.

\bibitem{van2020deterministic}
J.~van~den Brand, ``A deterministic linear program solver in current matrix multiplication time,'' in \emph{Proceedings of the Fourteenth Annual ACM-SIAM Symposium on Discrete Algorithms}.\hskip 1em plus 0.5em minus 0.4em\relax SIAM, 2020, pp. 259--278.

\bibitem{burke1995gauss}
J.~V. Burke and M.~C. Ferris, ``A {G}auss--{N}ewton method for convex composite optimization,'' \emph{Mathematical Programming}, vol.~71, no.~2, pp. 179--194, 1995.

\bibitem{clarke1990optimization}
F.~Clarke, \emph{Optimization and Nonsmooth Analysis}, ser. Classics in Applied Mathematics.\hskip 1em plus 0.5em minus 0.4em\relax Society for Industrial and Applied Mathematics, 1990.

\bibitem{netrapalli2013phase}
P.~Netrapalli, P.~Jain, and S.~Sanghavi, ``Phase retrieval using alternating minimization,'' \emph{Advances in Neural Information Processing Systems}, vol.~26, 2013.

\bibitem{parikh2014block}
N.~Parikh and S.~Boyd, ``Block splitting for distributed optimization,'' \emph{Mathematical Programming Computation}, vol.~6, no.~1, pp. 77--102, 2014.

\bibitem{holmstrom2009user}
K.~Holmstr{\"o}m, A.~O. G{\"o}ran, and M.~M. Edvall, ``User’s guide for tomlab/cplex v12. 1,'' \emph{Tomlab Optim. Retrieved}, vol.~1, p. 2017, 2009.

\bibitem{gurobi2021gurobi}
L.~Gurobi~Optimization, ``Gurobi optimizer reference manual,'' 2021.

\bibitem{yang2018rsg}
T.~Yang and Q.~Lin, ``Rsg: Beating subgradient method without smoothness and strong convexity,'' \emph{The Journal of Machine Learning Research}, vol.~19, no.~1, pp. 236--268, 2018.

\bibitem{ye1989extension}
Y.~Ye and E.~Tse, ``An extension of karmarkar's projective algorithm for convex quadratic programming,'' \emph{Mathematical programming}, vol.~44, pp. 157--179, 1989.

\bibitem{tan2019phase}
Y.~S. Tan and R.~Vershynin, ``Phase retrieval via randomized kaczmarz: theoretical guarantees,'' \emph{Information and Inference: A Journal of the IMA}, vol.~8, no.~1, pp. 97--123, 2019.

\bibitem{plan2014dimension}
Y.~Plan and R.~Vershynin, ``Dimension reduction by random hyperplane tessellations,'' \emph{Discrete \& Computational Geometry}, vol.~51, no.~2, pp. 438--461, 2014.

\bibitem{plan2012robust}
------, ``Robust 1-bit compressed sensing and sparse logistic regression: A convex programming approach,'' \emph{IEEE Transactions on Information Theory}, vol.~59, no.~1, pp. 482--494, 2012.

\end{thebibliography}

\end{document}